\documentstyle[12pt,epsf]{article}
\def\lbra       {\left( }
\def\rbra       {\right) }
\def\lmbra      {\left\{ }
\def\rmbra      {\right\} }
\def\lbbra      {\left[ }
\def\rbbra      {\right] }
\def\comma      { \, , }
\def\period     { \, . }
\def\bra#1      { \langle \, #1 \, \vert \, }
\def\ket#1      { \, \vert \, #1 \, \rangle \, }
\def\del        {  \partial  }
\def\half       {  {1\over 2}  }

\def\defint#1#2 {  \int_{#1}^{#2}  }
\def\rootof#1   {  \left( #1 \right)^{1/2}  }
\def\deldel#1   {  {\partial\over \partial #1}  }

\def\abs#1      {  \, \vert #1 \vert \,   }

\def\evalat#1   {  \left\vert_{#1} \right. }

\def\when       { \biggm{\vert} }
\def\lsim    {\lower .65ex \hbox{\ $\stackrel{<}{\sim}$\ } }
\def\gsim    {\lower .65ex \hbox{\ $\stackrel{>}{\sim}$\ } }

\def\calO       { {\cal O} }

\def\vecii#1#2      {  \left(\begin{array}{c}#1\\#2\end{array}\right)  }
\def\veciii#1#2#3   {  \left(\begin{array}{c}#1\\#2\\#3\end{array}\right)  }
\def\veciv#1#2#3#4  {  \left(\begin{array}{c}#1\\#2\\#3\\#4
                                 \end{array}\right)  }
\def\vecv#1#2#3#4#5 {  \left(\begin{array}{c}#1\\#2\\#3\\#4\\#5
                                 \end{array}\right)  }

\def\matrixii#1#2#3#4            {  \left(\begin{array}{cc}#1&#2\\#3&#4
                                       \end{array}\right) }
\def\matrixiii#1#2#3#4#5#6#7#8#9 {  \left(\begin{array}{ccc}#1&#2&#3\\
                                     #4&#5&#6\\#7&#8&#9\end{array}\right)  }
\def\mativ#1#2#3#4               {  \left(\begin{array}{cccc}
                                       #1\\#2\\#3\\#4\end{array}\right) }
\def\matv#1#2#3#4#5              {  \left(\begin{array}{ccccc}
                                     #1\\#2\\#3\\#4\\#5\end{array}\right)  }
\def\eqabegin         {  \begin{eqnarray}  }
\def\eqaend           {  \end{eqnarray}  }
\def\nn               {  \nonumber  }
\def\bracetwo#1#2     {  \left\{ \begin{array}{l} #1 \\ #2 \end{array}
                         \right.  }
\def\bracetwocases#1#2#3#4  {   \left\{ \begin{array}{ll} #1 &
                                 \qquad #2 \\
                                 #3 & \qquad #4 \end{array} \right.  }
\def\bracebegin#1     {  \left\{ \begin{array}{#1}   }
\def\braceend         {  \end{array}\right.   }
\def\parn              {  \par\noindent }
\def\parbigskip        {  \par\bigskip  }
\def\parmedskip        {  \par\medskip  }

\def\parbigskipn        {  \par\bigskip\noindent  }

\def\parag#1           {\paragraph{#1} \mbox{ }\parmedskip\noindent}

\def\splus   {\!+\!}
\def\sminus  {\!-\!}
\def\boxit#1#2      {  \vbox{\hrule\hbox{ \hskip -4.1pt \vrule\kern3pt
                       \vbox
                    {  \hsize #1 \strut\kern3pt #2 \kern3pt\strut  }
                       \kern3pt  \vrule} \hrule  } }
\def\centerbox#1#2  {  \mbox{  }\par\bigskip  \hfil \boxit{#1}{#2} \hfil
                       \par\bigskip\noindent }
\def\rightbox#1#2   {  \hfill\boxit{#1}{#2}  }
\def\leftbox#1#2    {  \boxit{#1}{#2}  }
\def\fullbox#1      {  \boxit{\textwidth}{#1}  }
\def\trianglemap#1#2#3#4#5#6  {   {\large $$ \begin{array}{rcl} #1\!\!\!
                                  &{\stackrel{{\scriptstyle #2}}{
                              \longrightarrow   }}&\!\!\!  #3 \\
                            { } & {\scriptstyle #4}\!\!\!\searrow \quad
                                \swarrow \!\!\!{\scriptstyle #5}& { } \\
                                  { } & #6 & { } \end{array} $$ }    }
\def\squaremap#1#2#3#4#5#6#7#8    { {\large $$ \begin{array}{ccc}#1 &
                   \stackrel{{\scriptstyle #2}}{\longrightarrow} & #3 \\
                     {\scriptstyle #4}\!\downarrow & { } & \downarrow \!
                     {\scriptstyle #5}\\ #6 &\!\!
                      \longrightarrow_{{ }_{\!\!\!\!\!\!\!\!\!\!\!
                      {\scriptstyle #7}}}    &#8 \end{array} $$ }   }
\def\righttrianglemap#1#2#3#4#5#6  {  {\large $$ \begin{array}{rcl}
                 #1\!\! & \stackrel{{\scriptstyle #2}}{\longrightarrow}
                      & #3 \\  { }&\!\!{\scriptstyle #4}\!\!\searrow
                      & \downarrow \!\!{\scriptstyle #5}\\
                      { }&{ }& #6 \end{array} $$ }   }
\def\rightfigspacebegin  {  \par\noindent\begin{minipage}[t]{10cm}  }
\def\rightfigspaceend    {  \end{minipage}\par\noindent  }
\def\leftfigspacebegin   {  \par\noindent
                             \hspace*{10cm}\begin{minipage}[t]{6cm} }
\def\leftfigspaceend     {  \end{minipage}\par\noindent  }
\def\titleandfile#1#2   {  \begin{center}{\Large\bf #1}\end{center}
                            \par\begin{flushright} #2 \end{flushright}  }
\def\sectionnumbering { \setcounter{equation}{0}
         \renewcommand{\theequation}{\arabic{section}.\arabic{equation}}}
\def\appendixnumbering#1 { \setcounter{equation}{0}
         \renewcommand{\theequation}{\arabic{#1}.\arabic{equation}}}
\def\msection#1      {  \begin{center} \section{#1} \end{center}   }
\def\nsection#1      {  \let\boldface\bf \def\bf{} \section{#1}
                           \let\bf\boldface   }
\def\mnsection#1     {  \begin{center} \nsection{#1} \end{center}  }
\def\csection#1    { \begin{center} \let\boldface\bf \def\bf{\large\sc}
               \section{#1} \let\bf\boldface \end{center} \sectionnumbering }
\def\csectionast#1    { \begin{center} \let\boldface\bf \def\bf{\large\sc}
               \section*{#1} \let\bf\boldface \end{center} \sectionnumbering }
\def\csubsection#1    {  \let\boldface\bf \def\bf{\normalsize\sc}
\subsection{#1}
                           \let\bf\boldface   }
\def\mcapsection#1   {  \begin{center} \csection{#1} \end{center} }
\newcommand{\nullify}[1]{}
\def\tilr        {\tilde{r}}
\def\tilt        {\tilde{t}}
\def\tilphi      {\tilde{\phi}}
\def\rp          {r_+}
\def\rmi         {r_-}
\def\vp          {v_+}
\def\vm          {v_-}
\def\psion       {\psi_{n \omega}}
\def\fon         {f_{n \omega}}
\def\gon         {g_{n \omega}}
\def\omh         {\Omega_H}
\def\th0         {\theta_0}
\def\Gf          {G_F}
\def\Gfe         {G_F^E}
\def\Gbh         {G_{BH}}
\def\Gbhe        {G_{BH}^E}
\def\tGf         { \tilde{G}_{F}^{E} }
\def\dH          {d_H}
\def\aH          {a_H}
\def\bH          {\beta_H}
\def\eH          {\epsilon_H}
\def\phip        {\phi^+}
\def\delt        { \Delta t }
\def\deltau      { \Delta \tau }
\def\delphi      { \Delta \phi }
\def\delvphi     { \Delta \varphi }
\def\delom       { s }
\def\tr              { \ {\rm tr } \ }
\def\cnp         { \ c_n^+ \ }
\def\cnm         { \ c_n^- \ }
\def\cn          { \ c_n \ }
\def\sn          { \ s_n \ }
\def\An          { A_n }
\def\Bn           { B  }
\def\bbh          {2 \pi \beta/\bH}

\def\papertitlepage{\baselineskip 3.5ex \thispagestyle{empty}}
\def\Title#1{\vspace{1.5cm}\begin{center}
 {\large\bf #1} \end{center}
\vspace{1.0cm}}
\def\Authors#1{\begin{center} {\sc #1} \end{center}}
\def\Address#1{\begin{center} {\rm #1} \end{center}}
\def\Abstract{\vspace{1.5cm}\begin{center} {\large\bf Abstract}
           \end{center} \parbigskip}
\def\Komabanumber#1#2#3{\hfill \begin{minipage}{4cm} UT-Komaba #1
              \parn #2\parn #3 \end{minipage}}
\renewcommand{\thefootnote}{\fnsymbol{footnote}}
\renewenvironment{thebibliography}{\pagebreak[3]\par\vspace{0.6em}
\begin{flushleft}{\large \bf References}\end{flushleft}
\vspace{-1.0em}

\begin{enumerate}\if@twocolumn\baselineskip=0.6em\itemsep -0.2em
\else\itemsep -0.2em\fi\labelsep 0.1em}{\end{enumerate}}
\begin{document}
%%%%%%%%%%%%%%%%%%%%%%%%%%%%
\papertitlepage
\vspace*{0cm}
\Komabanumber{94-23}{hepth/9412144}{December 1994}
%%%%%%%%%%%%%%%%%%%%%%%%%%%%
\Title{ENTROPIES OF SCALAR FIELDS \\ \vskip 1.5ex ON \\
\vskip 1.5ex THREE DIMENSIONAL BLACK HOLES }
\Authors{ Ikuo~Ichinose
\footnote[2]{e-mail address:\quad
ichinose@tansei.cc.u-tokyo.ac.jp}
 \  and \ \ Yuji~Satoh
\footnote[3]{e-mail address:\quad
ysatoh@hep1.c.u-tokyo.ac.jp} }
\vskip 2.8ex
\Address{
 Institute of Physics, University of Tokyo, \\
 Komaba, Tokyo 153 Japan
  }
%\vspace{1.5cm}
%%%%%%%%%%%%%%%%%%%%%%%%%%%%%%%%
\renewcommand{\thefootnote}{\arabic{footnote}}
%%%%%%%%%%%%%%%%%%%%%%%%%%%%%%%%%%%%%%%
\Abstract
%%%%%%%%%%%%%%%%%%%%%%%%%%%%%%%%%%%%%%%
\noindent
Thermodynamics of scalar fields is investigated in three dimensional
black hole backgrounds in two approaches. One is mode expansion and
direct computation of the partition sum, and the other is
the Euclidean path integral
approach. We obtain a number of exact results, for example,
 mode functions, Hartle-Hawking
Green functions on the black holes, Green functions on a cone geometry,
free energies and entropies. They constitute reliable bases for
the thermodynamics of scalar fields. It is shown that
thermodynamic quantities largely depend upon the approach to calculate
them, boundary conditions for the scalar fields and regularization method.
We find that, in general, the entropies are not
proportional to the area of the horizon and that their divergent
parts are not necessarily due to the existence of the horizon.
%
%%%%%%%%%%%%%%%%%%%%%%%%%%%%%%%%%%%%%%
\baselineskip=0.7cm
%%%%%%%%%%%%%%%%%%%%%%%%%%%%%%%%%%%%%
%\\
%\\
%PACS number(s): 04.06.+n
\newpage
%
%%%%%%%%%%%%%%%%%%%%%%%%%%%%%%%%%%%%%%%%%%%%%%
\baselineskip=0.7cm
%%%%%%%%%%%%%%%%%%%%%%%%%%%%%%%%%%%%%%%%%%%
%%%%%%%%%%%%%
\csection{Introduction}
%%%%%%%%%%%%%
Thermodynamics of black holes has been an enigma in theoretical physics.
It stands just at the junction of general relativity, quantum mechanics and
statistical mechanics. We have thought that its understanding leads
to physics beyond that we have at present, in particular, quantum gravity.

However,
we have not yet understood the true meaning of thermodynamic laws of
black holes,
and neither important related problems such as Hawking radiation and
quantum coherence.

In respect of these issues, some proposals have been made recently
\cite{SU}-\cite{Sol}.
For example, (i) the Bekenstein-Hawking entropy and the entropy of
a quantum field in a black hole background are the same object, i.e., the
response of the Euclidean path integral to the introduction of a conical
singularity to the underlying geometry \cite{GH,CW}.
(ii) The entropy of the quantum
field is obtained by tracing over local degrees of freedom inside the horizon
( `` geometric entropy '' ) \cite{Sr,CW}, by explicit counting of states
\cite{SU}
or by the Euclidean path integral.
(iii) It is proportional to the area of the horizon and gives the first
quantum correction to the Bekenstein-Hawking entropy \cite{SU,CW}.
(iv) Divergences
appear due to the blow up of density of states associated to
the horizon \cite{tH,SU,CW}
 and they can
be removed by the renormalization of the gravitational coupling
constant \cite{SU}.
(v) Consequently, the problem of information loss and that of
renormalizability of quantum gravity are intimately related.

Nevertheless, the system of a four dimensional black hole and a scalar field
is quite complicated. Thus we have to resort to some approximations and
somewhat formal arguments. Moreover, we do not know
whether the various
approaches for calculating the thermodynamic quantities are equivalent.
There is no reason that the equivalence must holds a priori.
These prevent us from clear understanding of the
arguments.

In this article, we shall work with the three dimensional black holes
of Einstein gravity discovered by
Ba${\rm \tilde{n}}$ados, Teitelboim and Zanelli \cite{BTZ,BHTZ}.
The three dimensional black holes share
many of the features of those in four dimensions, and moreover
they provide us with considerably simple systems.
Thus we can expect to avoid technical difficulties in four dimensions
and to be able to perform explicit analysis of their thermodynamics
including matter.

Therefore, we shall mainly pursue two purposes in this article.
One is to construct
reliable bases for the thermodynamics of quantum scalar fields in the three
dimensional black hole backgrounds. The other is to clarify the validity
of the recent arguments explained above by explicit calculations.
We believe that our results serve for deep understanding of thermodynamics
of black holes.

We organize the rest of this article as follows.
First, we briefly review the three dimensional black holes in Sec.2 .
Next, in Sec.3 , we study
the statistical mechanics of quantum scalar fields
by explicit mode expansion and direct computation of the partition sum.
We consider two boundary
conditions in order to examine the dependence of the thermodynamic quantities
upon boundary conditions .
One requires the regularity of the scalar fields at the
origin. In this case, we obtain exact expressions of the thermodynamic
quantities at an arbitrary temperature such as the free energies
and the entropies. The other requires the regularity of the scalar fields
at the outer horizon with a cutoff \cite{tH}. We have explicit forms of the
thermodynamic quantities and estimate them in the limit of the vanishing
cutoff.
In order to examine the equivalence between various approaches to the
thermodynamics of quantum fields, we study it also in the
Euclidean path integral approach. For this purpose,
Sec.4 is devoted to construction of Green functions of scalar
fields with arbitrary mass on the three dimensional black holes. It
turns out that our construction gives the Green functions
defined with respect to the Hartle-Hawking vacuum \cite{HH,GP}.
By making use of the Euclidean Hartle-Harking Green functions, we investigate
the statistical mechanics of the scalar fields in Sec.5 . We obtain
exact forms of free energies at the Hawking temperature.
 Then, we construct the Green functions with arbitrary period with respect to
the imaginary time, namely, those on a cone geometry. They enable us to
obtain exactly the thermodynamic quantities at an arbitrary temperature.
 In particular,
we calculate the entropies at the Hawking temperature and estimate their
divergent parts.
Finally, in Sec.6 , conclusions and discussions are given.
The construction
of Green functions in the universal covering space of
three dimensional anti-de Sitter space (CAdS${}_3$), which is necessary
in Sec.4, is summarized in Appendix.
%
%
%%%%%%%%%%%%%%%%%%%%%%%%%%%%
%%%%%%%%%%%%%%%%%%%%
\csection{ Three Dimensional Black Holes }
%%%%%%%%%%%%%%%%%%%%
Let us begin with a brief review of the three dimensional
  black hole discovered by
Ba${\rm \tilde{n}}$ados et al \cite{BTZ,BHTZ}. The three dimensional
 black hole is most easily obtained by
making use of some identifications under a discrete subgroup of the isometry
group of three dimensional anti-de Sitter space (AdS${}_3 $).

AdS${}_3$ is realized as the three dimensional hyperboloid
\eqabegin
           - u^2 - v^2  + x^2 + y^2 &=& - l^2
         \comma
\eqaend
in a four dimensional space with the metric
\eqabegin
      ds^2 &=& -du^2 -dv^2 + dx^2 + dy^2 \period
\eqaend
We introduce two parameters $ \rp $, and $ \rmi $ $ ( \rp \geq \rmi ) $
, which turn out shortly
to be the radii of the outer and inner horizons of the black hole,
and
we perform the following
 transformation into coordinates $ ( t, r, \phi ) $ :
\begin{equation}
   \begin{array}{lll}
    {\rm Region \ I.} \quad \rp < r  \quad :
   & u = \sqrt{\tilr^2} \cosh \tilphi \comma &
          v = \sqrt{\tilr^2 - l^2} \sinh \tilt \comma \nn \\
   &x = \sqrt{\tilr^2} \sinh \tilphi \comma &
          y = \sqrt{\tilr^2 - l^2} \cosh \tilt \comma  \label{reI} \\
   && \nn \\
   {\rm Region \ II.} \quad \rmi < r  < \rp \quad :
   & u = \sqrt{\tilr^2} \cosh \tilphi \comma &
          v = \sqrt{l^2 - \tilr^2 } \cosh \tilt \comma \nn \\
   &  x = \sqrt{\tilr^2} \sinh \tilphi \comma &
          y = \sqrt{l^2- \tilr^2 } \sinh \tilt \comma  \\
   && \nn \\
   {\rm Region \ III.} \quad 0 < r  < \rmi \quad :
   &u = \sqrt{-\tilr^2} \sinh \tilphi \comma &
          v = \sqrt{l^2 - \tilr^2 } \cosh \tilt \comma \nn \\
   & x = \sqrt{-\tilr^2} \cosh \tilphi \comma &
          y = \sqrt{l^2- \tilr^2 } \sinh \tilt \comma
   \end{array}
\end{equation}
where
\eqabegin
\tilr^2 = l^2 \left( \frac{r^2 - \rmi^2}{\rp^2 - \rmi^2} \right) \comma &&
\vecii{\tilt}{\tilphi} = \frac{1}{l}
  \matrixii{\rp}{-\rmi}{-\rmi}{\rp} \vecii{t/l}{\phi}
\period
\eqaend
With the above coordinates, the metric becomes
\eqabegin
 ds^2 &=& - N^2 dt^2 + N^{-2} dr^2 + r^2 (N^{\phi} dt + d\phi)^2 \comma
          \nn \\
      &=& -\left[ \frac{r^2}{l^2} - M \right] dt^2 - J dt d\phi +
           \left[ \frac{r^2}{l^2} - M + \frac{J^2}{4r^2} \right]^{-1} dr^2
           + r^2 d\phi^2
           \comma
\eqaend
with $ - \infty < t, \phi < + \infty $.
Here
\eqabegin
   N^2 \, = \, \frac{(r^2 - \rp^2)(r^2 - \rmi^2)}{l^2 r^2 } \comma &&
   N^{\phi} \, = \,  - \frac{\rp\rmi}{l r^2} \comma \\
   l^2 M \, = \, \rp^2 + \rmi^2 \comma && l J \, = \, 2 \rp\rmi \comma
\eqaend
and $ M $ and $ J $ are identified with the mass and the angular momentum
of the black hole, respectively.

The metric has a Killing vector $ \del_\phi $. Then by making the
identifications under the discrete subgroup of the isometry
group generated by
this Killing vector,
\eqabegin
 \phi & \longrightarrow & \phi + 2 \pi n \comma \quad ( n \in {\bf Z} )
  \comma
\eqaend
we get the black hole spacetime.

Note that the scalar curvature is a constant,
\eqabegin
 R &=& -6 l^{-2} \comma
\eqaend
because the black hole spacetime is locally AdS${}_3 $.

In the rest of the present paper, we shall consider quantum scalar
fields in the three dimensional black hole backgrounds and their
thermodynamics.

%\end{document}
%%%%%%%%%%%%%%%%%%%%%%%%%
%\input{notesty}
%\input{lmydef}
%\input{bhdefs}
%\begin{document}
%%%%%%%%%%%%%%%%%%%%
\csection{ Statistical Mechanics of Scalar Fields : Partition Sum}
%%%%%%%%%%%%%%%%%%%%%%
In this section, we consider the thermodynamics of scalar fields by
mode expansion and direct computation of the partition sum. In this
approach, the relation between the entropy and state counting is clear.
In order to study dependence of the thermodynamic quantities upon
boundary conditions, we consider two cases. In both cases, we require
that the scalar fields tend to vanish rapidly enough at spatial infinity.
In addition, we impose on them regularity at the origin for one case, or
at the horizon for the other case. The former is usually adopted
for wave functions, and the latter is adopted in the so-called ``
brick wall '' model \cite{tH,SU}.
Although it is possible to consider other various boundary conditions, we
do not take them because physical meaning is not clear in most cases.
It turn out that the thermodynamic
quantities largely depend upon the boundary conditions.

%%%%%%%%%%%%%%%%%%%%%%
\parn
\csubsection{Mode functions}
%%%%%%%%%%%%%%%%%%%%%%
Now let us consider a scalar field with mass squared $ m^2 $
in the three dimensional
 black hole back ground.
The field equation is given by
\eqabegin
 ( \Box - \mu l^{-2} ) \psi(x) &=& 0 \period
\eqaend
Since $ R = -6 l^{-2} $,
$ \mu l^{-2} = m^2 $ for a scalar field minimally coupled to
the background metric and
$ \mu l^{-2} = m^2 + (1/8) R = m^2 - (3/4)l^{-2} $
for a conformally coupled scalar
field.
In terms of the coordinates $ (t, r, \phi) $, the D'Alembertian operator,
$ \Box $, is given by
\eqabegin
 \Box \psi &=& \frac{1}{\sqrt{-g}} \del_a \left( \sqrt{-g} g^{ab} \del_b
               \right) \psi \nn \\
           &=& - \frac{1}{r^2 N^2} \left[ r^2 \del_t^2
                 - \left( \frac{r^2}{l^2} - M \right) \del_\phi^2 + J \del_t
                 \ \del_\phi \right] \psi + \frac{1}{r} \del_r \
                 \left( r N^2 \del_r \psi \right)
            \period
 \eqaend
 Changing variables to $ v = r^2/l^2 $, the field equation
 becomes as
 \eqabegin
  0 &=& \left[ v l^2  \del_t^2 - (v-M) \del_\phi^2 + J \del_t \ \del_\phi
            + \mu \Delta(v) \right] \psi - 4 \Delta(v) \del_v
            \left( \Delta(v) \del_v \psi \right)
            \comma
 \eqaend
 where
 \eqabegin
  \Delta(v) &=& \left( v - \frac{\rp^2}{l^2} \right)
                \left(v - \frac{\rmi^2}{l^2} \right) \, \equiv \,
                ( v - \vp )( v - \vm ) \period
 \eqaend
 The above equation is solved through separation of variables :
 \eqabegin
  \psion &=& \ e^{ - i \omega t } \ e^{i n \phi} \fon (v)
  \comma
 \eqaend
 where $ n $ is an integer.
 The equation for the radial function is given by
 \eqabegin
  \fon^{''} + \frac{\Delta'(v)}{\Delta(v)} \fon' + \frac{1}{4 \Delta^2(v)}
  \left\{ n (Mn - J \omega) - \mu \Delta(v) - (n^2-l^2\omega^2) v  \right\}
  \fon &=& 0 \comma \label{radeq}
 \eqaend
 where we have denoted the derivative with respect to $ v $ by the prime.

This equation has three regular singular points at
$ v = \vm, \vp, \infty $ corresponding to the inner horizon, the outer horizon
and the spatial infinity, respectively. Thus the solution is given
by hypergeometric functions. This is confirmed as follows.
First, let $ \fon $ be of the form
\eqabegin
 \fon &=& ( v - \vp )^\alpha (v - \vm )^\beta  \gon
 \comma
\eqaend
where $ \alpha $ and $ \beta $ are purely imaginary numbers defined by
\eqabegin
  \alpha^2 &=& - \frac{1}{4 (\vp -\vm)^2}
  \left( \rp \omega - \frac{\rmi}{l} n
              \right)^2 \comma \nn \\
  \beta^2 &=& - \frac{1}{4 (\vp -\vm)^2}
    \left( \rmi \omega - \frac{\rp}{l} n
    \right)^2  \label{albet} \period
\eqaend
The signs of $ \alpha $ and $ \beta $ are irrelevant in the following
discussion and we do not bother to specify them. Let us make the change
of variables
\eqabegin
  u &=& \frac{1}{\vp-\vm}( v- \vm) \period
\eqaend
Then Eq.(\ref{radeq}) becomes
\eqabegin
 u(1-u) \gon''(u) + \left\{c- (a+b+1)u \right\} \gon'(u) - a \, b\, \gon (u)
&=& 0
 \comma
\eqaend
where
\eqabegin
  a &=& ( \alpha + \beta ) + \half ( 1 + \sqrt{1 + \mu} ) \comma \nn \\
  b &=& ( \alpha + \beta ) + \half ( 1 - \sqrt{1 + \mu} ) \comma \nn \\
  c &=& 2 \beta + 1 \period \label{abc}
\eqaend
This is nothing but the hypergeometric equation, and $ \gon(u) $ is given by
hypergeometric functions.

The hypergeometric equation has two independent solutions around each regular
singular point. Thus we must impose boundary conditions
to specify the solution. First, we consider the behavior
of $ \fon(v) $ as $ v \to \infty \quad ( u \to \infty ) $.
Near the infinity, we have two independent solutions :
\eqabegin
 \fon^{1,\infty}(v) &=& (v-\vp)^\alpha (v-\vm)^\beta u^{-a}
          F( a, a \sminus c \splus 1; a \sminus b \splus 1; 1/u ) \comma \\
 \fon^{2,\infty}(v) &=& (v-\vp)^\alpha (v-\vm)^\beta u^{-b}
          F( b, b \sminus c \splus 1; b \sminus a \splus 1; 1/u ) \comma
\eqaend
where $ F $ is the hypergeometric function. From (\ref{abc}), we find that
$ \fon^{2,\infty} $  becomes divergent as $ v \to \infty $ for $ \mu > 0 $,
 while $ \fon^{1,\infty} $
comes to vanish for arbitrary $ \mu $.

The authors of \cite{AIS}-\cite{MT}
have discussed  the quantization of scalar fields in anti-de Sitter
space or its covering space, which has timelike spatial infinity
and needs special boundary conditions there.
If we require the condition to conserve energy following them,
the surface integral
of the energy momentum tensor
$ \lim_{r \to \infty} \int d S_i \sqrt{-g} T^i_{\ \ t} $
must vanish. This means
$ \sqrt{r} \fon \to 0 \quad ( r \to \infty )$, and only $ \fon^{1,\infty} $
satisfies this condition. Therefore we concentrate on $ \fon^{1,\infty} $
and drop the superscript $ ( 1, \infty ) $ in the following.

%%%%%%%%%%%%%%%%%%%%%%
\csubsection{Case I : regularity at the origin}
%%%%%%%%%%%%%%%%%%%%%%

In this section, we impose on $ \fon(v) $ regularity
at the origin $ ( r = 0 ) $ as is usual for radial functions,
and study the thermodynamics under this boundary condition.
Introducing appropriate cutoffs, we obtain exact results.

It is easy to see that $ \fon $
is regular at the origin because $ r = 0 $ corresponds to none of $ z = 0, 1,
\infty $. Thus we have no restriction on the value of $ \omega $.
Then we proceed to calculate thermodynamic quantities. First,
we consider the case of $ J \neq 0 $.
Recall that
the system of the rotating black hole and the scalar field has a chemical
potential $ \omh $. This is the angular velocity of the outer horizon:
\eqabegin
  \omh &=& \frac{d \phi}{d t } \biggm{\vert}_{r = r_+}
        \, = \, - N^{\phi} \biggm{\vert}_{r = r_+} \, = \, \frac{\rmi}{l\rp}
        \label{omh} \period
\eqaend
In addition, the system has superradiant scattering modes given by the
condition
\eqabegin
  \omega - \omh \ n \leq 0 \comma \label{srmode}
\eqaend
where $ \omega $ and $ n $ are the energy and angular momentum of the
scalar field, respectively.
Thus we have to regularize the ( grand ) partition function
by introducing the cutoff $ N_1 $ for the occupation number of
the particle for
each mode satisfying  (\ref{srmode}).

With these remarks in mind, we get the partition function for a single mode
labeled by $ \omega $ and $ n $,
\eqabegin
 Z_o( \beta; \omega, n ) &=& \sum_{m = 0}^{\infty}
   \ e^{-m ( \omega - \omh n )}
    \nn \\
    &=& \left\{ \begin{array}{cl}
                  \left( 1- \ e^{-\beta(\omega - \omh n )} \right)^{-1}
                 & {\rm for} \quad \omega - \omh n > 0 \\
                  N_1
                 & {\rm for} \quad \omega - \omh n = 0 \\
              %   & \\
                  \frac{1 - \ e^{- N_1 \beta(\omega - \omh n )}}
                  {1 - \ e^{- \beta(\omega - \omh n )}}
                 & {\rm for} \quad \omega - \omh n < 0
                \end{array} \right.
         \period
\eqaend
Then we obtain the total partition function,
\eqabegin
 Z_o(\beta) &=& \prod_{\omega, n} \ Z_o(\beta; \omega, n) \comma
\eqaend
and the free energy,
\eqabegin
 - \beta F_o(\beta) &=& \sum_{\omega, n} \ \ln \ Z_o(\beta; \omega, n ) \nn \\
                  &=& - \sum_{ \abs{n} = 0 }^{N_2} \frac{1}{\delom}
               \int_0^{\infty} d \omega
                   \ln \left( 1- \ e^{-\beta(\omega - \omh n )} \right)
                    + \sum_{n=0}^{N_2} N_1 \nn \\
                  && \qquad
                  + \sum_{n = 0}^{N_2} \frac{1}{\delom}
                   \int_0^{n \omh} d \omega
                   \ln \left( 1- \ e^{-\beta N_1 (\omega - \omh n )} \right)
                   \comma
 \eqaend
 where $ N_2 $ is  the cutoff for
 the absolute value of quantum number $ n $, and $ \delom $ is
the minimum spacing of $ \omega $. Note that
 $ \delom^{-1} $ is the density of states and
the above result is divergent as $ s \to 0 $
regardless of the existence of the horizon.
By making the change of variables $ t = \beta (\omega - \omh n ) $ for the
 first term and $ t = N_1\beta ( n \omh - \omega ) $ for the third term
 , we obtain
 \eqabegin
  - \beta F_o(\beta)
  %    &= &
  %        \frac{1}{(\delom)  \beta} \zeta (2)
  %     \left(\sum_{ \abs{n} = 0}^{N_2} \right)
  %       + N_1 \left(\sum_{n=0}^{N_2} \right) \nn \\
  %      && \qquad   + \frac{1}{s} \sum_{n = 1}^{N_2}
  %        \left[ \frac{\beta}{2} \omh^2 (N_1-1) \ n^2 + \frac{1}{N_1 \beta}
  %        \int_0^{N_1 \beta \omh n} d t \ \ln \left( 1 - \ e^{-t} \right)
  %        \right] \nn \\
      & =  & \frac{1}{\delom} \lbbra
        \frac{\pi^2}{6 \beta} ( 2 N_2 + 1)
     + \frac{\beta}{12} \omh^2 (N_1 \!-\!1) N_2 (N_2 \!+ \!1)(2N_2 \!+\!1)
             \right. \nn \\
       &&   \qquad  \left.
                  + \frac{1}{N_1 \beta} \sum_{n=1}^{N_2}
          \int_0^{N_1 \beta \omh n} d t \ \ln \left( 1 - \ e^{-t} \right)
          \rbbra
        + N_1 ( N_2 +1 )
      \period
\eqaend
%where $ \zeta(z) $ is the Riemann zeta function.
In the limit
$ N_1 \to \infty $, the last term in the bracket is simplified to be
$ - N_2 \zeta(2)/(N_1 \beta) $.

Since entropy is given by
\eqabegin
 S(\beta) &=& \beta^2 \frac{\del F }{\del \beta}
 \comma
\eqaend
we get
\eqabegin
  S_o(\beta) &=& \frac{1}{\delom} \lbbra
       \frac{\pi^2}{3 \beta} ( 2 N_2 + 1)
     - \omh \sum_{n = 1}^{N_2} n \ \ln \left( 1 - \ e^{-N_1 \beta \omh n }
       \right)
             \right. \label{SoJ} \\
       &&   \qquad  \left.
                  + \frac{2}{N_1 \beta} \sum_{n=1}^{N_2}
          \int_0^{N_1 \beta \omh n} d t \ \ln \left( 1 - \ e^{-t} \right)
          \rbbra
        + N_1 ( N_2 +1 )
     \nn \period
\eqaend

For the $ J = 0 $ case, the chemical potential vanishes,
and the partition function for a single mode
is given by
\eqabegin
 Z_o( \beta; \omega, n ) &=& \sum_{m = 0}^{\infty}
   \ e^{-m \omega }
    \nn \\
    &=& \left\{ \begin{array}{cl}
                  \left( 1- \ e^{-\beta \omega } \right)^{-1}
                 & {\rm for} \quad \omega > 0 \\
                  N_1
                 & {\rm for} \quad \omega = 0 \\
                 \end{array} \right.
         \period
\eqaend
Then the free energy becomes
\eqabegin
 - \beta F_o(\beta) &=& \sum_{\omega, n} \ \ln \ Z_o(\beta; \omega, n ) \nn \\
   &=& \frac{(2 N_2  \splus 1) \pi^2}{6 \delom \beta} + N_1 ( 2 N_2 + 1 )
  \period
\eqaend
Finally, we get the entropy ;
\eqabegin
    S_o(\beta)&=& \frac{(2 N_2  \splus 1) \pi^2}{3 \delom  \beta}
                      + N_1 ( 2 N_2 + 1 )
   \period \label{So}
\eqaend

{}From the expressions of the entropies (\ref{SoJ}) and (\ref{So}),
we find that the entropies are not proportional to
the area of the outer horizon $ ( 2 \pi \rp ) $  and that their divergences
are not due to the existence of the outer horizon.

%%%%%%%%%%%%%%%%%%%%%%
\csubsection{Case II : regularity at the outer horizon}
%%%%%%%%%%%%%%%%%%%%%%

Since the redshift factor of the black hole becomes divergent
at the horizon, one may expect that something singular occurs there.
Indeed, we find that $ \fon $ becomes singular.
Thus another natural boundary condition is to require regularity at the outer
horizon \cite{tH,SU}.
In this section, we introduce a cutoff for the distance from the outer
horizon in order to regulate $ \fon $.
Then we impose the regularity at the outer horizon on the scalar fields,
and study the thermodynamics under this condition.
At present, we do not
know whether this is the unique boundary condition which is physically
acceptable, and leave this as an open problem.

First, let us study the behavior of $ \fon $ near the outer horizon
( $ r = \rp $, i.e., $ u = 1 $ ).
By making  use of a linear transformation formula with respect to the
hypergeometric function, we get
\eqabegin
  \fon(v) &\propto& ( u -1 )^\alpha
     u^\beta F(a,b; 2\alpha  \splus 1; 1 \sminus u )
          \nn \\
          && \qquad + \Theta \ (u-1)^{-\alpha} u^{-\beta}
          F(1 \sminus b,1 \sminus a; -2\alpha  \splus 1;1 \sminus u)
          \comma
\eqaend
where
\eqabegin
 \Theta &=&
   \frac{\Gamma(1 \sminus b)\Gamma(c \sminus b)\Gamma(a \splus b \sminus c)}
  {\Gamma(a)\Gamma(a \sminus c \splus 1)\Gamma(c \sminus a \sminus b)} \period
\eqaend
{}From $ (\Gamma(z) )^\ast = \Gamma(z^\ast) $ and Eq.(\ref{abc}), we see that
$ \abs{\Theta} = 1 $. Thus we can set
\eqabegin
 \Theta &=& - \ e^{-2\pi i \theta_0} \qquad ( 0 \leq \theta_0 < 1 )
 \comma
\eqaend
where $ \theta_0 $ is determined by $ \omega $ and $ n $ through $ a $, $ b $
and $ c $.
Thus by choosing an appropriate normalization constant, it follows that
\eqabegin
  \fon(v) &=& ( u  \sminus 1 )^\alpha u^\beta \ e^{\pi i \theta_0}
          F(a,b; 2\alpha  \splus 1; 1 \sminus u )
          \nn \\
          && \qquad - (u \sminus 1)^{-\alpha} u^{-\beta} \ e^{-\pi i \theta_0}
           \ F(1 \sminus b,1 \sminus a; -2\alpha  \splus 1;1 \sminus u)
          \period
\eqaend

Then by introducing an infinitesimal constant, $ \eH $, and
substituting $ u = 1 + \eH^2/l^2 $ ( $ \eH^2 \propto r^2 - \rp^2 $ )
into $ \fon $ ,
we get the behavior
of $ \fon(v) $ near the outer horizon as follows
\eqabegin
 \fon &\stackrel{\eH \to 0 }{\longrightarrow}&
      \ e^{\alpha \ln (\eH^2/l^2 ) + \pi i \th0 } -
      \ e^{- (\alpha \ln ( \eH^2/l^2 ) + \pi i \th0 ) } \period
\eqaend
Clearly, the radial function is singular at the outer horizon.
Therefore we require
\footnote{
In \cite{SU}, the Dirichlet boundary condition at the horizon is
imposed on a scalar field. This is given by
$ K_{i \omega} (\xi \epsilon) = 0 $
where $ K_{i \omega} $ is the modified Bessel
function, $ \xi $ is a function of the momentum and the mass,
$ \xi = \sqrt{\vec{k}^2 + m^2} $, and $ \epsilon $ is essentially the same
as $ \eH $.
Supposed that $ \xi \epsilon << 1 $,
then the boundary condition is solved
by expanding $ K_{i \omega} $  to be $ \omega \ln (\xi \epsilon/2) \sim
 k \pi $.
This confirms the result by the WKB method adopted in \cite{SU} in the
leading order of $ \epsilon $.
}
that $ \fon $ vanishes at $ u = 1 + \eH^2/l^2 $
instead of $ u =1 $ .

Since $ \alpha $ is purely imaginary, the condition is easily realized.
{}From (\ref{albet}), we get
\eqabegin
 \omega &=& \omh \ n + C
               ( k + \th0 ) \comma \qquad  ( k \in {\bf Z} ) \comma \\
 C &=& \frac{2(\vp  \sminus \vm) \pi}{\rp \ln(l^2/\eH^2)}
               \period
\eqaend
This shows that $ \omega $ and $ \theta_0  $ are labeled by two integers
$ k $ and $ n $, i.e., $ \omega = \omega(k,n) $,
$ \theta_0 = \theta_0(k,n)$. Note that $ C^{-1} $ becomes singular as
$ \eH \to 0 $.

Let us consider thermodynamic quantities for
 $ J \neq 0 $ first.
They are obtained in the same way as in the previous
section. The partition function for a single mode labeled by $ k $
and $ n $ is given by
\eqabegin
 Z_h( \beta; k, n ) &=& \sum_{m = 0}^{\infty}  \ e^{-m ( \omega - \omh n )}
    \nn \\
    &=& \left\{ \begin{array}{cl}
                  \left( 1- \ e^{-\beta(\omega - \omh n )} \right)^{-1}
                 & {\rm for} \quad k + \th0 > 0 \\
                  N_1
                 & {\rm for} \quad k + \th0 = 0 \\
               % & \\
                  \frac{1 - \ e^{- N_1 \beta(\omega - \omh n )}}
                  {1 - \ e^{- \beta(\omega - \omh n )}}
                 & {\rm for} \quad k + \th0 < 0
                \end{array} \right.
         \period
\eqaend
Hence we get the total partition function,
\eqabegin
 Z_h(\beta) &=& \prod_{k, n} \ Z_h(\beta; k, n) \period
\eqaend
and the free energy,
\eqabegin
 - \beta F_h(\beta) &=& \sum_{k, n} \ \ln \ Z_h(\beta; k, n ) \nn \\
                  &=& - \sum_{ \abs{n} = 0 }^{N_2}
                  \sum_{ { k+ \theta_0 \neq 0 }
                  \atop{ C(k + \theta_0) \geq - \omh  n }}
                   \ln \left( 1- \ e^{-C \beta(k + \theta_0)} \right)
                    + N_1 \sum_{n=0}^{N_2} \delta_{\theta_0(0, n), 0} \nn \\
                  && \qquad
                  + \sum_{n = 1}^{N_2} \quad
                  \sum_{ 0>  C(k + \theta_0) \geq - \omh n}
                   \ln \left( 1- \ e^{- N_1 C \beta (k + \theta_0)} \right)
                   \period
 \eqaend
Since $ C<<1 $ in the limit $ \eH \to 0 $,
the summation with respect to $ k $ can be approximated
by integrals.
First, note that
\eqabegin
 \frac{d k }{d \omega} &=& \frac{1}{C}
             - \frac{d \theta_0}{d k }\frac{d k }{d \omega}
          \ \sim \ \frac{1}{C}
       \period
\eqaend
This shows that the density of states diverges due to the existence of
the outer horizon
i.e., as $ \eH \to 0 $.
Then in the same way as
in the previous section,
we get
 \eqabegin
  - \beta F_h(\beta) & \sim &
             \frac{\rp l^2 \delom \ln (l^2/\eH^2)}{2\pi \dH^2}
                       \lmbra -\beta F_o(\beta) -  N_1(N_2 \splus 1) \rmbra
                       +
                       N_1 \sum_{n=0}^{N_2} \delta_{\theta_0(0,n), 0}
                \comma
\eqaend
where
\eqabegin
  \dH^2 &=& \rp^2 - \rmi^2 \period
\eqaend

As for the entropy, we have
\eqabegin
 S_h(\beta) &\sim&
    \frac{\rp l^2 \delom \ln (l^2/\eH^2)}{2\pi \dH^2}
                       \lmbra  S_o(\beta) - N_1(N_2 \splus 1) \rmbra
                     +
                       N_1 \sum_{n=0}^{N_2} \delta_{\theta_0(0,n), 0}
                \period  \label{ShJ}
\eqaend

In the case of $ J = 0 $, the calculation is performed in the same way,
and we obtain
\eqabegin
 - \beta F_h(\beta) & \sim &
         \frac{\pi \rp l^2 \ln (l^2/\eH^2)}{12 \dH^2  \beta}(2N_2 \splus 1)
                + N_1 \sum_{\abs{n} = 0}^{N_2} \delta_{\theta_0(0,n), 0}
       \comma \\
 S_h(\beta) &\sim&
         \frac{\pi \rp l^2 \ln (l^2/\eH^2)}{6 \dH^2 \beta} (2N_2 \splus 1)
                + N_1 \sum_{\abs{n} = 0}^{N_2} \delta_{\theta_0(0,n), 0}
       \period \label{Sh}
\eqaend

{}From the expressions of the entropies (\ref{ShJ}) and (\ref{Sh}), we find
(i) that the leading terms of the entropies as $ \eH \to 0 $ are proportional
to $ \rp $ , but there can
exist the terms which is not proportional to $ \rp $,
(ii) that the entropies diverges due to the outer horizon as $ \eH \to 0 $
like $ \ln (l^2/\eH^2) $.

%\end{document}
%%%%%%%%%%%%%%%%%%%%%%%%
%\input{notesty}
%\input{lmydef}
%\input{bhdefs}
%\begin{document}
%%%%%%%%%%%%%%%%%%%%
\csection{ Green Functions on the Three Dimensional Black Holes}
%%%%%%%%%%%%%%%%%%%%
In the preceding section, we calculated thermodynamic
quantities by the straightforward mode sum.
In the following, we shall investigate thermodynamics of scalar fields
in the Euclidean path integral approach
in order to examine the equivalence between the various approaches.
For this purpose, we construct Green functions of scalar fields on the
black hole and identify the vacuum state in this section.

%%%%%%%%%%%%%%%%%%%%%%
\parn
\csubsection{ Construction of green functions}
%%%%%%%%%%%%%%%%%%%%%%

Quantization  of a scalar field in the universal covering space of
$ n$-dimensional anti-de Sitter space (CAdS${}_n $) is discussed in
\cite{AIS}-\cite{BL}, and the Feynman Green function is given
in terms of hypergeometric functions \cite{BL}.
In the three dimensional case $ ( n=3 ) $,
by using a mathematical formula for hypergeometric functions,
we get fairly simple form of the Green function,
\eqabegin
 - i \Gf (x,x') &=& - i \Gf(z) \ \equiv \
           \frac{1}{4 \pi l} ( z^2 - 1)^{-1/2}
           \lbbra z + ( z^2 - 1 )^{1/2} \rbbra^{1-\lambda} \label{Gf} \comma
\eqaend
where
\eqabegin
 z &=& 1 + l^{-2} \sigma (x, x') + i \varepsilon \comma \\
 && \nn \\
 \lambda &=&  \lmbra
           \begin{array}{ll}
              \lambda_\pm \, \equiv \, 1 \pm \sqrt{1 + \mu}
              & {\rm for } \ \ 0 > \mu > -1 \\
              & \\
              \lambda_+  & {\rm for} \ \     \mu \geq 0 \comma \mu = -1
           \end{array}
              \right.
              \comma
\eqaend
( two $ \lambda $'s are possible for $ 0 > \mu > -1 $ ).
$ \sigma(x,x') $ is half of the distance between $ x $ and $ x'$ in the four
dimensional embedding space,
\eqabegin
  \sigma (x, x') &=& \half \eta_{\mu\nu} ( \xi - \xi')^{\mu}( \xi - \xi')^{\nu}
              \comma
\eqaend
where $ \eta_{\mu\nu} = diag( -1,-1, +1,+1 )$ and $ \xi $ and $ \xi' $ are
the coordinates in the embedding space.
Since the derivation of this result is somewhat lengthy and technical,
we relegate it to Appendix.

By making use of the above result,
Green functions in the three dimensional
 black hole background are obtained
by the method of images \cite{SM1}-\cite{St} ;
\eqabegin
 - i \Gbh (x,x') &=& - i \sum_{n = - \infty}^{\infty} \Gf (x, x'_n)
                         \nn \\
    &=& \frac{1}{4 \pi l} \sum_{n = - \infty}^{\infty} ( z_n^2 - 1)^{-1/2}
           \lbbra z_n + ( z_n^2 - 1 )^{1/2} \rbbra^{1-\lambda}
           \label{Gbh}\comma
\eqaend
where
\eqabegin
  x_n &\equiv& x \ \Bigm{\vert}_{\phi'\to\phi'- 2n \pi} \comma \qquad
  z_n(x, x') \, = \, z(x, x'_n)
  \period
\eqaend
By using (\ref{reI}), we have
\eqabegin
z_n(x,x') - i \varepsilon &=& \frac{1}{\dH^2}
         \lbbra \sqrt{r^2 -\rmi^2} \sqrt{r'^2 - \rmi^2}
         \cosh \lbra \frac{\rmi}{l^2} \Delta t - \frac{\rp}{l} \Delta \phi_n
               \rbra \right.
               \nn \\
      & & \left. \qquad \qquad -
          \sqrt{r^2 -\rp^2} \sqrt{r'^2 - \rp^2}
         \cosh \lbra \frac{\rp}{l^2} \Delta t - \frac{\rmi}{l} \Delta \phi_n
               \rbra \rbbra
               \label{zn} \comma
\eqaend
where
\eqabegin
% \dH^2 &=& \rp^2 - \rmi^2 \comma \\
 \Delta t &=& t - t' \comma \qquad \Delta \phi_n \, = \, \phi - \phi' + 2 n \pi
 \label{deltphi}
 \period
\eqaend

Note that in the case of the conformally coupled massless scalar field, namely
$ \mu = -3/4 $, $ \lambda_\pm = 3/2, 1/2 $ holds and the
corresponding Green functions
become
\eqabegin
 -i \Gbh(x,x') &=& \frac{1}{2^{\lambda + 1} \pi l} \sum_{n = - \infty}^{\infty}
             \lbbra \frac{1}{\sqrt{z_n -1}} \pm \frac{1}{\sqrt{z_n + 1}}
             \rbbra
             \period
\eqaend
The above results coincide with those of \cite{SM1}-\cite{St}
constructed from the Green functions with the ``Neumann"
or ``Dirichlet" boundary conditions in CAdS${}_3$.

%%%%%%%%%%%%%%%%%%%%%%
\csubsection{ Boundary conditions and the vacuum }
%%%%%%%%%%%%%%%%%%%%%%

In the previous section, we have constructed Green functions
on three dimensional
 black holes. However, the physical meaning of the Green functions
is not clear unless we specify its boundary conditions
and identify the vacuum with respect to which they are defined.

It turns out that $ \Gbh $ which we have constructed
satisfies the boundary conditions :
(i) to be regular at infinity (ii) to be analytic in the lower half plane
on the future complexified outer horizon (iii) to be analytic in the upper
half plane on the past complexified outer horizon. These conditions
fix $ \Gbh $ as a solution of the inhomogeneous wave equation
\cite{HH}. This means that $ \Gbh $ is regarded as the Green function
constructed by the Kruskal modes, namely as the Hartle-Hawking Green function.
In other words, the vacuum with respect to which $ \Gbh $ is defined is
identified with the Hartle-Hawking vacuum \cite{HH,GP}.

Now let us prove the above statement.
For brevity, we concentrate
only on the case $ r, r' \geq \rp $ in the following. It is easy to see that
the boundary condition (i) is satisfied from the definition of $ \Gbh $.

The conditions (ii) and (iii) have already been verified
for the case of the conformally coupled
massless scalar field on the non-rotating
black hole ( i.e. $ J = 0 $ ) \cite{LO}.
Thus we follow the strategy of \cite{LO}.

First, we introduce Kruskal coordinates \cite{BHTZ} by
\eqabegin
 V &=& R(r) \ e^{\aH t} \comma \qquad U \, = \, - R(r) \ e^{-\aH t}
   \comma \nn \\
 R(r) &=& \sqrt{\lbra \frac{r-\rp}{r + \rp} \rbra
     \lbra \frac{r + \rmi}{r -\rmi}\rbra^{\rmi/\rp}}
     \label{defKruskal}\comma
\eqaend
where
\eqabegin
   \aH &=& \frac{\rp^2 - \rmi^2}{\rp l^2} \period
\eqaend
In these coordinates, the metric becomes
\eqabegin
 d s^2 &=& \Omega^2(r) d U d V + r^2 \lbra N^\phi d t + d \phi \rbra^2
  \comma \\
 \Omega^2(r) &=& \frac{(r^2 - \rmi^2 )(r + \rp)^2}{\aH^2 r^2 l^2}
                 \lbra \frac{r-\rmi}{r + \rp}\rbra^{\rmi/\rp}
 \period
\eqaend

Second, let us recall the Kerr black hole. In this case, we have to
introduce an angle coordinate rotating with the outer horizon in order to
obtain the expression of the metric regular on the outer horizon and
to extend the spacetime maximally \cite{Cart}.
In the same way, we introduce a new
angle coordinate rotating with the outer horizon,
\eqabegin
 \phip &=& \phi - \omh t \comma \label{phip}
\eqaend
because we are interested in the situation just on the outer horizon.
Note that $ N^{\phi} d t + d \phi = d \phip $ on the outer horizon.
In this coordinate, it follows that
\eqabegin
z_n(x,x') - i \varepsilon &=& \frac{1}{\dH^2}
         \lbbra \sqrt{r^2 -\rmi^2} \sqrt{r'^2 - \rmi^2}
         \cosh \lbra \frac{\rp}{l} \Delta \phip_n
               \rbra \right.
               \nn \\
      & & \left. \qquad \qquad -
          \sqrt{r^2 -\rp^2} \sqrt{r'^2 - \rp^2}
         \cosh \lbra \aH \Delta t - \frac{\rmi}{l} \Delta \phip_n
               \rbra \rbbra
               \comma \label{znp}
\eqaend
where
\eqabegin
  \Delta \phi^+_n &=&= \phip - \phi'^+ + 2 n \pi  \period
%  \aH &=& \frac{\rp^2 - \rmi^2}{\rp l^2} \period
\eqaend

Then let us examine the analyticity of the Green functions on the
past complexified outer horizon given by $ V = 0 $ and Re$(-U) > 0 $.
In terms of $ (t, r) $, this condition is equivalent to
\eqabegin
 \lmbra
   \begin{array}{l}
    r \longrightarrow \rp \\
    t \longrightarrow - \infty \comma
   \end{array}
 \right.
   &{\rm with} & \sqrt{r- \rp} \ e^{-\aH t} \longrightarrow \sqrt{l} A \comma
  \label{pastho}
\eqaend
where $ A $ is a constant determined by the value of $ U $ and with
the property Re $ A > 0 $. It follows from (\ref{defKruskal})
that Im $A >0 $ and Im $A < 0 $
correspond to the lower
and the upper half planes of $ U $, respectively.

In the limit (\ref{pastho}), $ z_n $ becomes
\eqabegin
z_n(x,x') - i \varepsilon &\longrightarrow& \frac{1}{\dH^2}
         \lbbra \sqrt{\rp2 -\rmi^2} \sqrt{r'^2 - \rmi^2}
         \cosh \lbra \frac{\rp}{l} \Delta \phip_n
               \rbra \right.
               \nn \\
      & & \left. \qquad \qquad -
          \half \sqrt{2 \rp l} \sqrt{r'^2 - \rp^2}
         \ e^{ \aH t' + (\rmi/l) \Delta \phip_n } \ A \rbbra
                \period \label{znlim}
\eqaend
Recalling the form of $ \Gbh $, i.e. (\ref{Gbh}),
we see that each component
in the summation in $ \Gbh $ has singularities where $ z_n = \pm 1 $.
{}From $ r' > \rp $ and (\ref{znlim}), we see that the solutions
to $ z_n = \pm 1 $ on the past complexified outer horizon are of the
form
\eqabegin
 A &=& \alpha_0 + i \varepsilon \comma
\eqaend
for both signs, where $ \alpha_0 $ is some positive number.
Thus we find that the each component in $ \Gbh $ is regular
in the upper half plane of $ U $.

Then we shall use Weierstrass's theorem \cite{Ah} : if a series
with analytic terms converges uniformly on every compact subset of a region,
then the sum is analytic in that region, and the series can
be differentiated
term by term. In fact, we can check that
the series in $ \Gbh $ is converges uniformly.
Thus we conclude that
$ \Gbh $ is analytic in the upper half plan of the past complexified outer
horizon.

The proof of the analyticity on the lower half plan of the future complexified
 outer horizon is much the same. Thus we omit it for brevity.

Finally, we make a comment. The Hartle-Hawking Green function on a
black hole was originally defined in the path-integral formalism
as a generalization of the Feynman Green function in Minkowski
spacetime \cite{HH}. In our case, $ \Gf $ is also defined so as to
conform to the Feynman Green function in the flat limit
( see (\ref{epGf}) and the comment below it ). Thus it is natural that
$ \Gf $ satisfies the Hartle-Hawking boundary condition.

%\end{document}
%%%%%%%%%%%%%%%%%%%%%%%
%\input{notesty}
%\input{lmydef}
%\input{bhdefs}
%\begin{document}
%%%%%%%%%%%%%%%%%%%%
\csection{ Statistical Mechanics of Scalar Fields by Hartle -Hawking
Green Functions}
%%%%%%%%%%%%%%%%%%%
In the previous section, we see that the Green function $ \Gbh $ is the
Hartle-Hawking Green function, which is often used for discussion on
thermodynamics of black holes and Hawking radiation. In this section,
we discuss statistical mechanics of a scalar field by using $ \Gbh $.
In the Euclidean path integral approach to statistical field system,
thermodynamic quantities are obtained by making use of Euclidean Green
functions.

%%%%%%%%%%%%%%%%%%%%%%
\parn
\csubsection{ Euclidean green functions }
%%%%%%%%%%%%%%%%%%%%%%

Let us define the Green function on the Euclidean black hole
geometry. Introducing the Euclidean time $ \tau = i t $ and the
`` Euclidean '' angle $ \varphi = - i \phi $ for $ J \neq 0 $ and
$ \varphi = \phi $ for $ J = 0 $, this is given by
\eqabegin
  \Gbhe ( \deltau, \delvphi; r, r') &\equiv&
       \sum_{n = - \infty}^{\infty}  \Gfe ( \deltau, \delvphi_n; r, r')
       \comma \\
   \Gfe ( \deltau, \delvphi ; r, r') &\equiv&
     \lmbra
         \begin{array}{ll}
            i \Gf ( \delt, \delphi; r, r')
             \vert_{{\delt=i\deltau}\atop{\delvphi = \delphi}}
             & {\rm for} \quad J= 0 \\
             & \\
             i \Gf ( \delt, \delphi; r, r')
             \vert_{{\delt=i\deltau}\atop{\delvphi = - i \delphi}}
              & {\rm for} \quad J \neq  0
         \end{array}
          \right.
   \period
\eqaend
Here $ \deltau $ and $ \delvphi_n $ is defined as (\ref{deltphi}),
and the superscript $ E $ means Euclidean quantities.
The factors in front of $ \Gf $ are
chosen so that the physical quantities calculated later will
have real values with
appropriate signs.

Since $ ( \Box - \mu l^{-2} ) \Gbh =  (1/\sqrt{-g}) \delta(x-x') $,
we have
\eqabegin
 ( \Box^E - \mu l^{-2} ) \Gbhe &=&
                \lmbra
              \begin{array}{ll}
              - \frac{1}{\sqrt{ \abs{g^E} }} \ \delta^E (x-x')
              & {\rm for } \ J = 0 \\
              & \\
               \frac{i}{\sqrt{ \abs{g^E} }} \ \delta^E (x-x')
               & {\rm for } \ J \neq 0
             \end{array}
          \right.
          \label{eqGbhe} \comma
\eqaend
where $ g^E $
is defined by the line element
\eqabegin
 d s_E^2 &=&
          \lmbra
          \begin{array}{ll}
               N^2 d \tau^2  + r^2 d \varphi^2 + N^{-2} d r^2
                & {\rm for } \ J = 0 \\
                & \\
               N^2 d \tau^2  - r^2 ( N^\phi d \tau - d \varphi )^2
               + N^{-2} d r^2 & {\rm for } \ J \neq 0
          \end{array}
          \right. \period
\eqaend
Please note that the metric is not positive definite for $ J \neq 0 $.

Then let us consider thermal properties of $ \Gbhe $. For a moment,
we concentrate on the case of $ J \neq 0 $.
A thermal Green function at temperature
$ \beta^{-1} $ and with a chemical potential $ \nu $ conjugate to
angular momentum is defined by
\eqabegin
 G_{\beta}^E (x, x'; \nu) &=& {\rm tr} \
          \lbbra \ e^{ -\beta( \hat{H}- \nu \hat{L})} \ T
          \lbra \psi(x) \psi(x') \rbra \rbbra
          \biggm{/} {\rm tr} \
          \lbbra \ e^{ -\beta( \hat{H}- \nu \hat{L})} \rbbra
      \comma
\eqaend
where $ T $ denotes the ( Euclidean ) time ordered product and $ \hat{H} $ and
$ \hat{L} $ are
the generators of time translation and rotation, respectively.
{}From the above definition,
\eqabegin
  G_{\beta}^E (\tau, \varphi, r; \tau', \varphi', r'; \nu) &=&
  G_{\beta}^E
    (\tau + \beta , \varphi -  \nu \beta , r; \tau', \varphi', r'; \nu)
  \period
\eqaend

Because the Green function $ \Gbh $ is a function of $ z_n $, from (\ref{zn})
we find that $ \Gbhe $ is periodic under
\eqabegin
   \delta \lbra \frac{\rmi}{l^2} \  \tau + \frac{\rp}{l} \  \varphi \rbra &=&
      2 \pi m \  \nn \\
   \delta \lbra \frac{\rp}{l^2} \  \tau +  \frac{\rmi}{l} \  \varphi \rbra &=&
      2 \pi n \  \qquad
     ( m, n \in {\bf Z} ) \comma
\eqaend
where $ \delta ( ... ) $ means the variation of the arguments.
Thus $ \Gbhe $ is of double period
\eqabegin
  \vecii{\delta (\tau/l) }{ \delta \varphi} &=&
  \frac{2 \pi l }{\dH^2} \matrixii{-\rmi}{\rp}{\rp}{-\rmi}
   \vecii{m}{n}   \period
\eqaend
If we require that, as $ J \to 0 $ $ (\rmi \to 0 ) $,
the chemical potential vanishes,
the fundamental period is determined uniquely as
\eqabegin
  \tau  &\to & \tau + \frac{ 2 \pi }{\aH } \ n \comma \qquad
  \varphi  \, \to \, \varphi - \frac{ 2 \pi }{\aH l} \ \nu  \comma \\
 \nu &=& \frac{\rmi}{l \rp } \, = \, \omh
 \period
\eqaend
It is easy to see that this result is valid also in the case of $ J = 0 $.
Therefore, we conclude that $ \Gbhe $ can be regarded as
a thermal Green function at
the inverse temperature
\eqabegin
   \beta_H &\equiv & 2 \pi / \aH
   \comma
\eqaend
and with the chemical potential
$ \omh $. We shall calculate thermodynamic quantities by making use of
$ \Gbhe $.
In the following, we shall explicitly show the period of the Green
functions, for example,
as $ \Gbhe ( \deltau, \delvphi, r, r'; \bH ) $.

It is instructive to consider the behavior of the metric near the
outer horizon. Let us introduce a coordinate $ \eta $ by
\eqabegin
   r &=& \rp + \frac{2}{\aH} \eta^2
  \period
\eqaend
Then for small $ \eta $ , the metric becomes
\eqabegin
 d s^2 & \sim & - \aH^2 \ \eta^2 \  d t^2 + d \eta^2 + \rp^2 (d \phip)^2
   \label{metho} \period
\eqaend
Moreover, in terms of the Euclidean time $ \tau = - i t $ ,
$ ( \tau, r ) $ represents a plane with the origin $ r = \rp $.
Therefore we find that  $ \beta_H $ is  nothing
but the period around the outer
horizon of the Euclidean black hole, while $ \omh $ is the angular
velocity of the outer
horizon ( see (\ref{omh}) \ ).
Thus our result confirms
and gives an explicit example to
the arguments in the literature of thermodynamics of black holes.
In addition, it may be worth noting that for small $ \eta, \eta', \deltau,
\delvphi$,
\eqabegin
   2 \sigma(x,x') &\sim& ( \Delta \eta )^2 + \rp^2 (\delphi^+)^2
   \period
\eqaend
Thus the distance of the embedding space becomes
that with respect to the metric (\ref{metho}).

%Now we have establised the thermal property of $ \Gbh$.
%In the case at temperature $ \beta^{-1} $ and with the chemical potential
%$ \nu $, these are of period $ \tau \to \tau + \beta $ and
%$ \varphi \to \varphi - \nu \beta $. $ \tau $ and $ \varphi $ are
%imaginary time and angle: $ \tau = i t $ , $ \varphi = i \phi $.
%In our case, such  Green functions may be given by

%%%%%%%%%%%%%%%%%%%%%%
\csubsection{ Free energy }
%%%%%%%%%%%%%%%%%%%%%%

In this section, we shall calculate
the free energy $ F(\beta) $, which is given by
\eqabegin
   \beta F(\beta) &=&  - \half \tr \log \Gbhe (\beta)
             \label{trlog}  \comma
\eqaend
where the trace is defined by
\eqabegin
  \tr (\  ... \ ) &=&  \int d^3 x  \sqrt{\abs{g^E} } \lim_{x \to x'}
                   ( \ ... \ )
                     \nn \\
               &=&
          \lmbra
          \begin{array}{ll}
                   \int_0^{\beta} d \tau \int_0^{2 \pi} d \varphi
                     \int_{\rp}^{\infty} d r \cdot r \ \lim_{x \to x'}
                     ( \ ... \ )
                & {\rm for } \ J = 0 \\
                 & \\
                   \int_0^{\beta} d \tau \int_0^{\omh \beta}
                   d \varphi
                     \int_{\rp}^{\infty} d r \cdot r \ \lim_{x \to x'}
                     ( \ ... \ )
               & {\rm for } \ J \neq 0
          \end{array}
          \right. \label{deftr} \period
\eqaend
In (\ref{deftr}), we have set
the lower end of the integration with respect to $ r $ to be
$ \rp $. The reasons are twofold; (i) in the Euclidean geometry,
the topology of $ (\tau, r ) $
space is $ {\bf R}^2 $ and the origin corresponds to $ r = \rp $,
and (ii) it turns out that the entropy becomes complex if we perform
integration below $ \rp $.

{}From the expressions (\ref{eqGbhe}) and (\ref{trlog}), it follows that
\eqabegin
  \frac{\del}{\del \mu} \lbra \beta  \ F(\beta) \rbra
  &=& - \frac{1}{2 l^2} \tr \Gbhe (\beta)
  \label{trG} \period
\eqaend
In the case of flat spacetime, the expression like (\ref{trlog}) is
divergent. Thus we have to regularize it by the expression like (\ref{trG}).
For getting the right answer,
we then integrate out (\ref{trG}).
However since  the derivation from (\ref{trlog}) to (\ref{trG}) is
rather formal,
the final result may depend upon with which of these we start. Indeed, we
shall find
that the result depends on the choice.
Here we  start with (\ref{trG}) according to the prescription
in the case of flat spacetime.

We  consider the case $ J \neq 0 $ first. In this case, we have
\eqabegin
  \frac{\del}{\del \mu} \lbra \beta \ F(\beta) \rbra
          \biggm{\vert}_{\beta = \bH} &=&
          - i \ \frac{\omh}{4 l^2} \beta^2 \sum_{n = - \infty}^{\infty}
          \int_{\rp^2}^{\infty} d (r^2)
           \lim_{r \to r'}  \Gf ( z_n^0; \bH)
           \ \when_{\beta = \bH}
           \comma \label{delF}
\eqaend
where $ z_n^0 = z_n \when_{\deltau = \delvphi = 0 } $ and we have used
the fact that the integrand is independent of $ \tau $ and $ \varphi $.

Recalling the expression of $ \Gf $ and $ z_n $, i.e., (\ref{Gf}) and
(\ref{zn}),
we see that the integrand with $ n = 0 $ in the summation
diverges in the limit $ r  \to r ' $. Thus we remove this term
for a  moment.

It is useful to notice that $ \Gf(z_n; \bH) $ and $ z_n^0 $ are written
as
\eqabegin
  - i \Gf (z_n; \bH) &=&
          \lmbra
            \begin{array}{ll}
              \frac{l^{-1}}{4 \pi} \frac{1}{1 - \lambda} \frac{d}{dz_n}
              \ e^{(1-\lambda) \coth^{-1} z_n }
              & {\rm for } \ \lambda \neq 1 \\
              & \\
              \frac{l^{-1}}{4 \pi} \lbra z_n^2 -1 \rbra^{-1/2}
              & {\rm for } \ \lambda = 1 \\
            \end{array}
          \right. \comma \\
    && \nn \\
        z_n^0 \when_{r = r'} &=& \frac{1}{\dH^2}
            \lmbra (r^2 - \rmi^2) \cnp - (r^2 - \rp^2 ) \cnm \rmbra
            \comma
\eqaend
where
\eqabegin
 \cn^\pm &=& \cosh \lbra 2 \pi n \frac{r_\pm}{l} \rbra
  \period
\eqaend
Since the infinitesimal imaginary part of $ z_n $ is irrelevant for the
discussion, we have omitted it. We shall do it also in the following unless
it is necessary.
Then by making the change of variables from $ r^2 $ to $ z_n^0 $, we get
\eqabegin
 && -i \int_{\rp^2}^{\infty} d (r^2)  \lim_{r \to r'} \Gf(z_n^0; \bH) \nn \\
 && \qquad = \ \lmbra
      \begin{array}{ll}
        \frac{l^{-1}}{4\pi} \frac{\dH^2}{(\cnp - \cnm)(1 -\lambda)}
         \lbra  z + \sqrt{z^2-1} \rbra^{1 -\lambda} \when_{\cnp}^{\infty}
       & {\rm for } \lambda \neq 1 \\
       & \\
       \frac{l^{-1}}{4\pi} \frac{2 \dH^2}{(\cnp - \cnm)}
         \log \lbra  z + \sqrt{z^2-1} \rbra \when_{\cnp}^{\infty}
       & {\rm for } \lambda = 1
      \end{array}
     \right. \period
\eqaend
Therefore the integral diverges at the upper end for $ \lambda < 1 $
( i.e.,  $ \lambda = \lambda_- $ ) and for $ \lambda = 1 $
( i.e.,  $ \lambda = \lambda_+ \comma \  \mu = -1 $),
while for $ \lambda > 1 $ ( i.e., $ \lambda = \lambda_+ , \mu \neq -1 ) \ $
 we get
\eqabegin
 &&  \frac{\del}{\del \mu} \lbra \beta \ F(\beta) \rbra
          \biggm{\vert}_{\beta = \bH} \nn \\
 &&  \qquad \qquad =
  \frac{l^{-3}}{8 \pi} \omh \bH^2
  \frac{\dH^2}{\lambda-1} \sum_{n \geq 1}^{\infty} \frac{1}{\cnp - \cnm} \
   e^{ - 2 \pi (\lambda-1) n \rp/l} \ + C_0  \comma
\eqaend
where $ C_0 $ is the divergent term coming from $ n = 0 $.
By integrating the above expression, we obtain
\eqabegin
   \beta \ F(\beta)
          \biggm{\vert}_{\beta = \bH}
 &=&
  - \frac{l^{-2} \omh}{8 \pi^2 \rp}
     \bH^2 \dH^2 \sum_{n \geq 1}^{\infty}
    \frac{1}{n (\cnp -\cnm)}
    \
   e^{ - 2 \pi (\lambda-1) n \rp/l}  \nn \\
 & & \qquad \qquad \qquad \qquad + \ \ {\rm const.}
     \qquad \qquad ({\rm for} \ \lambda > 1 ) \period
\eqaend

For the case of $ J = 0 $, the calculation is performed
in a similar way, and the result is
\eqabegin
   \beta \ F(\beta)
          \biggm{\vert}_{\beta = \bH}
 &=&
  -   \frac{\dH^2 \bH}{4 \pi \rp  l^2}
    \sum_{n \geq 1}^{\infty} \frac{1}{n (\cnp - \cnm)}
    \
   e^{ - 2 \pi (\lambda-1) n \rp/l}  \nn \\
 & & \qquad \qquad \qquad + \ \ {\rm const.}
     \qquad \qquad ({\rm for} \ \lambda > 1 ) \period
\eqaend

%[ in the limit the sum -> int. ]

%%%%%%%%%%%%%%%%%%%%%%
\csubsection{ Green functions on a cone geometry }
%%%%%%%%%%%%%%%%%%%%%%

In order to calculate the entropy, we have to differentiate the
Euclidean Green function with respect to $ \beta $ with the chemical
potential fixed.
Thus we need Green functions with
period different from $ \beta_H = 2 \pi/\aH $ with $ \omh $ fixed.
Namely, we need to construct Green functions on $ \tau - r $ plane
with a deficit angle around the origin, i.e., on a cone geometry.
For this purpose, we first regard
$ \deltau $ and $\delvphi^+$ $ ( = \delvphi + \omh \deltau ) $
as independent variables, and then
we fix the value of $ \delvphi^+$.
After that, we construct the Green
function with an arbitrary period $  \beta $ with respect to $ \deltau $.
By this procedure, it is assured that the chemical potential is unchanged.

Long time ago,
the problem of constructing  solutions of certain differential equations
with period different from $ 2 \pi $ from the one with the period $ 2 \pi $
was discussed \cite{So,Ca}. Then this method
was applied to field theory on curved spaces \cite{Do,Fu1,Sol}.
We can also make use of this method to obtain the
Green function with an arbitrary period $ \beta $.

We introduce a new variable $ w $ by $ w = \aH \tau = - i \aH t $ and
denote $ \Gfe(z_n; \bH) $ with  $ \delvphi^+_n $, $ r $ and $ r' $ fixed
by
\eqabegin
    \tGf ( w - w'_n; 2 \pi ) &\equiv& \Gfe ( \deltau, \delvphi_n; r, r'; \bH )
                    \when_{ \delvphi^+, r, r' : \ {\rm fixed} }
    \comma
\eqaend
where
\eqabegin
  w_n &=& w - i \frac{\rmi}{l} \delphi^+_n
 \period
\eqaend
Note that $ \tGf $ depends upon $ w - w'_n $ through $ z_n ( x, x' ) $
as
\eqabegin
  z_n(x,x') - i \varepsilon &=& \frac{1}{\dH^2}
         \lbbra \sqrt{r^2 -\rmi^2} \sqrt{r'^2 - \rmi^2}
         \cosh \lbra \frac{\rp}{l} \Delta \phip_n
               \rbra \right.
               \nn \\
      & & \left. \qquad \qquad -
          \sqrt{r^2 -\rp^2} \sqrt{r'^2 - \rp'^2}
         \cosh \lbra i ( w - w'_n )
               \rbra \rbbra
               \period \label{zww}
\eqaend
Then we shall study the location of singularities of $ \tGf $.  This
is necessary for writing out the expression of the Green function
with an arbitrary period $\bbh$, i.e., $ \tGf(w - w'; \bbh) $.
In the same way as  $ \Gf(z_n; \bH) $, $ \tGf $ has singularities
at $ z_n =
\pm 1 $. From (\ref{zww}),
%this condition is reduced to
%\eqabegin
%   \cosh \lbra \aH \delt - \frac{\rmi}{l} \delphi^+_n \rbra  &=&
%   \frac{1}{\sqrt{r^2-\rp^2}\sqrt{r^{'2} - \rp^2}}
%   \lbra \sqrt{r^2 - \rmi^2} \sqrt{r^{'2} - \rmi^2 } \cnp \mp \dH^2 \rbra
%    \pm i \varepsilon
%   \comma \nn \\
% &&
%\eqaend
%for $ r, r' > \rp $.
we find that  $ \tGf $ has four singularities in the region
$ - \pi < $ Re $ (w - w'_n) \leq \pi $.
They are located infinitesimally close to the imaginary axis of
$ w - w'_n $ plane and symmetrically with respect
to the point
$ w - w'_n = 0 $.
In the limit
$ \delvphi^+_n \to 0 $ and $ r \to  r' $,  two of these singularities approach
to the same point $ w - w'_n
 = 0 $.

Now we are ready to construct $ \tGf( w-w'_n; \bbh )$.
This is given by the Sommerfeld integral representation \cite{So,Ca},
\eqabegin
  \tGf( w - w'_n; \bbh ) &=& \frac{\bH}{2 \pi \beta} \int_\Gamma d \zeta
  \tGf( \zeta - w'_n; 2 \pi )
  \frac{ e^{ i \bH \zeta/\beta}}
{ e^{ i \bH \zeta/\beta}- \ e^{ i \bH w /\beta}}
  \comma
\eqaend
where the contour $ \Gamma $ of the integral is given by the solid
line in Fig.1 .
This contour consists of two parts and divide the four singularities into
two pairs. In the case of $ \delvphi^+_n = 0 $ and $ r = r' $,  we cannot take
such a contour because two of the singularities degenerate into
$ \zeta - w'_n = 0 $ .
Therefore we define $ \tGf $ in this case by
\eqabegin
 \tGf (w- w'_n;\bbh) \when_{{\delvphi_n^+ = 0 \comma }\atop{r = r'}} &\equiv&
 \lim_{{\delvphi_n^+ \to 0 \comma }\atop{r \to r'}} \tGf (w-w'_n; \bbh)
 \when_{{\delvphi_n^+ \neq 0}\atop{{\rm or} \ r \neq r'}}
 \period
\eqaend
%In addition, when $ w - w'_n= \pi $ holds,
%there may be the singularity $ \zeta = w $ of the integrand just on
%the contour $ \Gamma $.
%Thus, in this case, we first regularize the expression
%by an infinitesimally positive number $ \varepsilon $ as
%$ w - w'_n = \pi - \varepsilon $. Then we take the limit
%$ \varepsilon \to 0 $.
{}From $ \tGf( w-w'_n; \bbh )$, we can get the Green functions with an
arbitrary period $ \beta $ on the cone geometry, namely,
$ \Gbhe(x-x'; \beta) $ in a similar way as (\ref{Gbh}) :
\eqabegin
   \Gbhe(x-x'; \beta) &=&
       \sum_{n = - \infty}^{\infty} \Gfe( x-x'_n; \beta)
       \comma \\
     \Gfe( x-x'_n; \beta) &=& \tGf ( w - w'_n; \bbh )
   \period
\eqaend

It is instructive to consider some special cases
before we prove that the expression above is actually correct.
First, in the case of $ \beta = \bH/q \comma \ (q = 1,2,...) $,
the contour $ \Gamma $ is deformed into $ \Gamma'$
given by the dashed line in Fig.1 .
Since the integrand is of period $ 2 \pi $,
the contributions from the path made
up of straight lines cancel with each other. Thus
only the residues inside the circular path contribute to the integral.
Therefore we get
\eqabegin
 \tGf (w -w'_n; 2 \pi/q) &=& \sum_{k} \tGf(w(k) - w'_n; 2 \pi)
   \label{2pq}
 \comma
\eqaend
where $ w(k) $ and $ k ( \in {\bf Z}) $ are given by
$ w(k) =  w + 2\pi k/q $ and
$ - \pi < w (k) \leq \pi $.
In this case, the method of images works and we can explicitly check the
periodicity. Clearly, in the case of $ q = 1 $, the r.h.s. of
(\ref{2pq}) reproduces $ \tGf(w-w'_n; 2 \pi) $.

Next, let us consider the case $ \beta \to \infty $.
In the limit $ \beta \to \infty $, it follows that
\eqabegin
   \tGf(w -w'_n: \infty) &=& \frac{1}{2 \pi i } \int_\Gamma
    \tGf(w -w'_n; 2 \pi)
    \frac{d \zeta}{\zeta - w }
    \period
\eqaend
By using this and a formula
$ \lim_{n \to \infty} \sum_{k = -n}^{n} 1/(x + k) = \pi \cot \pi x $,
we obtain another expression for $ \tGf(w-w'_n;\bH) $ :
\eqabegin
 \tGf(w-w'_n;\bbh)&=& \sum_{k = - \infty}^{\infty}
                            \tGf(w-w'_n + 2 \pi k \beta/\bH ; \infty) \nn \\
        &=&
        \frac{\bH}{ 4 \pi i \beta} \int_\Gamma  d \zeta
       \tGf(\zeta - w'_n; 2 \pi)
        \cot \lmbra \frac{\bH}{2 \beta}( \zeta - w ) \rmbra
        \period \label{cotform}
\eqaend
The equivalence to the former expression is easily checked by noting
$ \tGf(w-w'_n;\bH) = \tGf(w-w'_n;- \bH) $.

Now we shall check properties of Green functions.
First, it is clear that $ \tGf(w-w'_n;\bbh) $ actually converges  because
$ \tGf(\zeta -w'_n;\bbh) $ comes to vanish exponentially
as $ \abs{ {\rm Im} \ \zeta} \to \infty  $. It is also easy to see that
$ \tGf(w-w'_n;\bbh) $ is of period $ \bbh $ by making the change of variables
$ \zeta - w'_n = \zeta' $.

Finally, let us check that $ \Gfe(x-x'; \beta) $ satisfies
the inhomogeneous
equation. As an example, we consider the case of $ J = 0 $.
{}From
$ ( \Box_{x'}^E - \mu l^{-2} ) \Gfe(x-x'; \bH) =
                - ( 1/\sqrt{ \abs{g^E} }) $ $ \times \delta_{\bH}^E (x-x') $,
it follows
that
\eqabegin
( \Box_{x'}^E - \mu l^{-2} ) \Gfe(x-x'; \infty) &=&
                - \frac{1}{\sqrt{ \abs{g^E} }} \ \delta^E_{\infty} (x-x')
          \comma
\eqaend
where we have explicitly denoted the period of the delta function.
Thus
from the fact $ \Gfe(x-x'_n ;\beta) = \sum_{k = -\infty}^{\infty}
\Gfe(x-x'_n; \infty)\Bigm{\vert}_{ \deltau \to \deltau + k \beta} $ ,
we get the desired result :
\eqabegin
( \Box_{x'}^E - \mu l^{-2} ) \Gfe(x-x'; \beta) &=&
                - \frac{1}{\sqrt{ \abs{g^E} }} \ \delta^E_{\beta} (x-x')
          \period
\eqaend

For calculating entropy in the later section,
let us calculate the derivative of the Green functions.
{}From the cotangent form (\ref{cotform}), we get the derivative of
$ \Gfe(x-x'_n; \beta) $ with respect to $ \beta $ ,
\eqabegin
 && \frac{\del}{\del \beta} \Gfe(x-x'_n;\beta)
 = - \frac{1}{\beta} \Gfe(x-x'_n;\beta)  \label{delG} \\
  && \qquad \quad +
 \frac{\bH^2}{8 \pi i \beta^3} \int_\Gamma d \zeta \ \tGf(\zeta - w'_n; 2 \pi)
          ( \zeta - w )
        {\rm cosec}^2 \lmbra \frac{\bH}{2 \beta} (\zeta - w ) \rmbra \nn
   \period
\eqaend
%\frac{1}{2 i \beta} \int_\Gamma d \zeta \tGf(\zeta - w'_n; \bH) \times
% \nn \\
%&& \qquad \lbbra  -\frac{\bH}{2 \pi \beta} \cot \lmbra \frac{\bH}{2 \beta}
%   ( \zeta - w ) \rmbra + \frac{\bH^2 (\zeta-w)}{. 4 \pi \beta^2} \
%    {\rm cosec}^2 \lmbra \frac{\bH}{2 \beta} (\zeta - w ) \rmbra \rbbra
%    \period \label{delG}
%\eqaend
In the case of $ \beta = \bH $, the above expression is fairly simplified.
First, we deform the contour $ \Gamma $ into $ \Gamma' $.
In this case, the singularity
within the circular path is only at $ \zeta = w $, and the contribution
from the
residue of this singularity cancels with
the first term in (\ref{delG}).
%Moreover,
%the contributions from the path made up of two straight lines also cancel
%each other in the first term because the first term is of period
%$  2 \pi $.
Thus by changing variables, we get
\eqabegin
  \frac{\del}{\del \beta} \Gfe(x-x'_n;\beta)
&=& \frac{1}{4 \bH} \int_{- \infty}^{\infty} d \zeta' \
    \frac{\tGf( i \zeta' - \pi; \bH )}
             {\cos^2 \lmbra ( i \zeta' + w'_n - w )/2 \rmbra}
 \period
\eqaend
%where we have defined $ \zeta' $ by $ i \zeta = \zeta' $.
Note that $ \tGf( i \zeta' - \pi; \bH ) $ is a function of
\eqabegin
   z ( \zeta' )&\equiv & z_n(x,x') \when_{w-w'_n = i \zeta' - \pi}
   = \An + \Bn \cosh \zeta'
  \comma
\eqaend
where
\eqabegin
  \An &=& \frac{\sqrt{r^2 - \rmi^2}\sqrt{r'^2 - \rmi^2}}{\dH^2}
    \cosh \lbra \frac{\rp}{l} \delphi_n^+ \rbra
  \comma \qquad \Bn \, = \,
    \frac{\sqrt{r^2 - \rp^2}\sqrt{r'^2 - \rp^2}}{\dH^2}
    \period
\eqaend
Thus we have
\eqabegin
 \frac{d z }{d \zeta'} &=&
 \Bn \sinh \zeta'
 \ = \ \pm \sqrt{( z - \An )^2 - \Bn^2}
 \comma
\eqaend
and
\eqabegin
  \cos^2 \lmbra \half (i \zeta' + w'_n - w)) \rmbra &=&
  \half \lmbra  \cn  \frac{z-\An}{\Bn}
    \pm \sn \frac{\sqrt{( z - \An )^2 - \Bn^2}}{\Bn} \ + \ 1 \rmbra
    \period
\eqaend
$ \cn $ and $ \sn $ are given by
\eqabegin
 \cn &=& \cosh \lbra i( w-w'_n) \rbra
 \comma \qquad
 \sn = \sinh \lbra i (w-w'_n) \rbra
 \period
\eqaend
Therefore by making the further change of variables from $ \zeta' $ to $ z $,
we get the fairly simple form :
\eqabegin
  && \frac{\del}{\del \beta} \Gfe(x-x'_n; \beta) \when_{\beta = \bH}  \nn \\
  &=& - \frac{\Bn}{\bH} \int_{\An + \Bn}^{\infty} d z
   \ \Gfe (z; \bH) \frac{1}{\sqrt{( z \sminus \An )^2 \sminus \Bn^2}}
   \frac{\cn (z \sminus \An) \splus \Bn}{( z \sminus \An \splus \cn \Bn )^2}
  \period \label{delGf}
\eqaend

Note that  $ \Bn >0  $ for $ r > \rp $ and $ \Bn < 0 $
for $ r< \rp $ hold . Thus $ \An + \Bn $ can be less than unity
for $ r < \rp $, and
the Green function $ -i\Gfe(z; \bH) $  becomes complex.
Moreover, it turns out that the entropy also becomes complex due to
the contribution from this region.
This indicates that we should consider only the region $ r > \rp $ in
calculation of thermodynamic quantities as
we have so far done.
%%%%%%%%%%%%%%%%%%%%%%
\csubsection{ Entropy }
%%%%%%%%%%%%%%%%%%%%%%

Now we are ready to calculate the entropy.
%This is given by
%\eqabegin
%   S(\beta) &=& \beta^2 \frac{\del F}{\del \beta}
%   \period
%\eqaend
First, we consider the case of $ J \neq 0 $.
{}From equations like Eq.(\ref{delF}) and Eq.(\ref{delGf}),
it follows that
\eqabegin
  && \frac{\del}{\del \mu} S(\beta) \when_{\beta = \bH}
   = - \frac{i}{4 l^2} \omh \bH^2
  \sum_{n = - \infty}^{\infty} \
\int_{\rp^2}^{\infty} d(r^2) \
     \lim_{r \to r'}
        \Biggl[ \Gfe(x-x'_n; \bH)
     \\
  && \quad     \left.
     -  \Bn \int_{\An \splus \Bn}^{\infty}
     d z \
     \Gfe(z; \bH) \
     \frac{1}{\sqrt{( z \sminus \An )^2 \sminus \Bn^2}}
\frac{\cn (z \sminus \An)\splus \Bn}{( z \sminus \An \splus \cn \Bn )^2} \rbbra
       \when_{\deltau = \delvphi = 0}
 \period \nn
\eqaend
The first term is nothing but $ \del_{\mu}( \beta F(\bH) ) $.
Thus by integrating the above expression, we get
\eqabegin
  && S(\bH) = \bH F(\bH)
   +   \frac{\omh}{8 \pi l^3 } \ \bH^2 \
  \sum_{n = - \infty}^{\infty} \
 \int_{\rp^2}^{\infty} d (r^2)
  \ \lim_{r \to r'}
    \  \Biggl[ \Bn \int_{\An \splus \Bn}^{\infty}
     d z
 \label{SJ} \\
  && \qquad  \times \left.
     \frac{X^{1 \sminus \lambda} \lmbra 1 + (\lambda \sminus 1)
     \log X \rmbra}
      {\log^2 X \sqrt{z^2  \sminus 1}}
      \frac{1}{\sqrt{( z \sminus  \An )^2 \sminus \Bn^2}}
\frac{\cn (z \sminus \An)\splus \Bn}{( z \sminus \An \splus \cn \Bn )^2} \rbbra
       \when_{\deltau = \delvphi = 0}
   + \ c  \comma \nn
\eqaend
where
\eqabegin
 X &=& z + \sqrt{z^2 -1}
 \comma
\eqaend
and $ c $ is a constant independent of $ \mu $ and
is ignored in the flat case.

For the case of $ J = 0 $, the calculation is modified
due to the difference in the definition of the trace. However we can
get the entropy in the same way as
\eqabegin
  && S(\bH)
   =   \frac{1}{4 l^3 } \ \bH \
\sum_{n = - \infty}^{\infty} \
\int_{\rp}^{\infty} d (r^2)  \
 \lim_{r \to r'} \
     \Biggl[ \Bn \int_{\An \splus \Bn}^{\infty}
     d z
 \\
&& \quad     \left. \times
\frac{X^{1 \sminus \lambda} \lmbra 1 \splus (\lambda \sminus 1) \log X \rmbra}
      {\log^2 X \sqrt{z^2  \sminus 1}}
      \frac{1}{\sqrt{( z \sminus  \An )^2 \sminus  \Bn^2}}
\frac{\cn (z \sminus \An ) \sminus \Bn}
{( z \sminus  \An \splus  \cn \Bn )^2} \rbbra
       \when_{\deltau = \delvphi = 0}
   + \ c  \period \nn
\eqaend

As we have exact expressions of the entropies,
we can study the structure
of their divergences without any ambiguity.
At present, it is believed that the divergence of the entropy
of quantum fields on black holes is closely related to important
physical problems \cite{SU}-\cite{Sol}.
As divergent parts in the entropy for $ J = 0 $ can be easily
obtained from those for $ J \neq 0 $,
we concentrate on the latter case first.

%There are several sources of the divergences. First, the integral with respect
%to $ z $ in the second term in (\ref{SJ}) behaves near the upper end
%like
%\eqabegin
% &&  \int^{\infty} d z \ z ^{- \lambda -2}
%  \period
%\eqaend
%Thus for $ \lambda = \lambda_- $ and $ \sqrt{1 + \mu} \geq 2 $, this integral
%diverges at the upper end.

It is possible that the integral of the second term in (\ref{SJ})
diverges by the contribution from the region of large $ r $.
However, we do not know whether it occurs
unless we know the behavior of the integrand in the second term for
large $ r $. Thus we leave this as an open question. This possible
divergence is regarded as an infrared divergence.

Next, let us consider divergences which come from short distances.
For this purpose, we introduce an infinitesimal variable $ \rho $ and
an infinitesimal constant $ s $ by
\eqabegin
   \rho^2 &=& r^2 - \rp^2 \comma \qquad s^2 \, = \, r'^2 - r^2
   \period
\eqaend
In the limit $ \rho, \ s \to 0 $, we have
\eqabegin
  \An &\sim& \cnp \lbra 1 + \frac{\rho^2 + s^2/2}{\dH^2} \rbra
 \comma \qquad \Bn = \lbra \frac{\rho}{\dH^2} \rbra \sqrt{\rho^2 + s^2}
 \comma \\
  z^0 &\equiv& z(x,x') \when_{\delt = \delvphi = 0} \sim
   1 + \frac{1}{\dH^2}
   \lbra \rho^2 + \frac{s^2}{2}- \rho \sqrt{\rho^2 + s^2}
   \rbra \nn \\
   && \qquad \qquad \qquad \quad
   \sim  1 + \frac{s^4}{8(r^2-\rp^2)(r^2 - \rmi^2)}
    \qquad  {\rm for} \ \rho^2 >> s^2
    \period
\eqaend
Since it turns out that the divergences due to short distances come only
from the term with $ n = 0 $ in the summation in (\ref{SJ}), we focus on this
term, which is given by
\eqabegin
    I &\equiv & \frac{\omh}{8 \pi l^3 } \ \bH^2 \ \int_{\rp}^{\infty} d (r^2)
      \ \lim_{s \to 0 } \ B \int_{A_0 \splus B}^{\infty}
     d z  \
 \\
  && \quad  \times \
     \frac{X^{1 \sminus \lambda} \lmbra 1 + (\lambda \sminus 1)
     \log X \rmbra}
      {\log^2 X \sqrt{z^2  \sminus 1}}
      \frac{1}{\sqrt{( z \sminus  A_0 )^2 \sminus B^2}}
\frac{ (z \sminus A_0)\splus B}{( z \sminus A_0 \splus  B )^2}
       \when_{\deltau = \delvphi = 0}
     \nn \period
\eqaend
By introducing two new variables by
\eqabegin
  z' &\equiv & z - A_0 \comma \qquad
  \delta  \, = \, A_0 -1
  \sim \frac{1}{\dH^2}\lbra \rho^2 + \frac{s^2}{2} \rbra
  \comma
\eqaend
we get $ \log X \sim \sqrt{2(z'+ \delta)} $ up to
$ \calO (z', \delta) $.
Therefore we find the
contribution to the integral from the region of the short distances,
\eqabegin
 I &\sim&
 \frac{\omh}{8 \pi l^3 } \ \bH^2  \int_{\eH^2} d(\rho^2) B \int_{B} d z'
 \frac{1 \splus ( \lambda \sminus 1) \sqrt{2(z' \splus \delta )}}
 {( z' \splus \delta)(z' \splus B)
 \sqrt{z' \splus \delta}\sqrt{z'^2 \sminus B^2}} \nn \\
 &=&
 \frac{\omh}{8 \pi l^3 } \ \bH^2  \int_{\eH^2} d(\rho^2)  \ B \ \int^{1} d u
  \\
  && \qquad \times \
 \frac{1}{\lbra 1 \splus (\delta/B) u \rbra (1 \splus u)
  \sqrt{1 \sminus u^2}}  \
   \lmbra
  \frac{1}{\sqrt{1 \splus (\delta/B) \ u} } \ \lbra
  \frac{u}{B} \rbra^{3/2}
  + \sqrt{2} ( \lambda \sminus 1 ) \frac{u}{B}
 \rmbra
 \nn \comma
\eqaend
where $ u = B/z' $.
We have regularized the integral by introducing the cutoff, $ \eH $, for
 the lower end of the integral as
\eqabegin
 \rp^2  &\longrightarrow & \rp^2 + \eH^2
 \period
\eqaend
Thus, for $ \eH, \rho \simeq s $ or $ \eH, \rho >> s $,
since $ \delta/B \sim 1 $
and $ B \sim (\rho/\dH)^2 $ hold, we obtain
\eqabegin
I & \sim & \omh \bH^2 l^{-3} \int_{\eH^2} d ( \rho^2 )
        \lbra B^{-3/2} + \ c \ B^{-1} \rbra \nn \\
   & \sim &
    \omh \bH^2 l^{-3} \lbra \dH^3  \frac{1}{\eH} + \ c' \dH^2 \log \eH^2  \rbra
    \comma
\eqaend
where $ c $ and $ c' $ are constants.
On the other hand for $ \eH, \rho << s $, since
$ \delta/B_0 \sim s/\rho $
holds, we obtain
\eqabegin
I & \sim & \omh \bH^2 l^{-3} \int_{s^2} d ( \rho^2 )
        \lbra \delta^{-3/2} + \ c \ \delta^{-1} \rbra \nn \\
   & \sim &
    \omh \bH^2 l^{-3} \lbra \dH^3  \frac{1}{s}
    + \ c' \dH^2 \log s^2  \rbra
    \period
\eqaend
Therefore the divergences are given in terms of the larger
cutoff, i.e.,  max$ \{ \eH, s  \} $.

Finally, we study divergences coming from the first term of (\ref{SJ}),
namely, $ \bH F(\bH) $. As discussed in Sec 5.2 , this term has two sources of
divergences. One is the integration over large $ r $  for $ \lambda \leq 1 $.
The other is the term with $ n = 0 $ in the summation in the integrand.
This term becomes divergent for small $ s $ like
\eqabegin
 && \frac{1}{\sqrt{\sigma(x,x')}} \sim \frac{1}{s^2}
    \sqrt{(r^2 - \rp^2)(r^2 - \rmi^2)}
    \period
\eqaend

For the case of $ J = 0 $, the analysis of divergent parts is performed
in a similar way. Thus we do not give details of it, and just
make some remarks.
In this case, the term corresponding to $ \bH F(\bH) $
and divergences associated with this term do not exist.
The entropy of the case $ J = 0 $ is obtained from
the second term in (\ref{SJ})
by replacing the factor $ \omh \bH $ with $ 2 \pi $. Thus
divergences of the entropy are easily obtained by the same procedure.

Finally we make some comments on the entropies.
First, from the expression of the entropies,
we find that they contain various divergences
coming from short
distances such as $ \eH^{-1}, \log \eH^2, $ and $ s^{-2} $. However
not all of them
are due to the existence of the outer horizon.
We also find that
the divergent terms are proportional to $ \dH $.
Hence they are proportional to the area of the outer horizon $ (\rp ) $
for the $ J = 0 $ case, but this does not hold for the $ J \neq 0 $ case.
Actually, the divergent terms vanish in the extreme limit $ \rmi \to \rp $.
This agrees with the discussion ( at the classical level )
that extreme black holes have zero entropy
\cite{HHR,Tei}.
%\end{document}
%%%%%%%%%%%%%%%%%%%%%
%%%%%%%%%%%%%%%%%%%%
\csection{ Conclusions and Discussions }
%%%%%%%%%%%%%%%%%%%%%%

We have investigated the thermodynamics of scalar fields
on the three dimensional black holes in two approaches. One ( approach I )
is based on
explicit mode expansion of the scalar fields and direct computation
of the partition sum,
and the other ( approach II ) is
based on Hartle-Hawking Green functions. In both approaches,
explicit expressions of the free energies and the entropies are obtained.
We believe that we have provided reliable bases for the study of
thermodynamics of
scalar fields on the
three dimensional black holes and that our results give useful insight for
understanding of thermodynamics of black holes in four dimensions.

Our results also allow us to answer
the interesting questions listed in the introduction at least
for the three dimensional case.

First, we obtained physical quantities such as densities
of states, free energies and entropies in approach I. Then we found that
they crucially depend upon boundary conditions.
In particular, divergent terms of the entropy are not necessarily due to
the existence of the outer horizon.
They also depend upon the boundary conditions.
In addition, the
entropy is not proportional to the area of the outer horizon, namely,
the Bekenstein-Hawking formula is invalid.

Second, we constructed exact Hartle-Hawking Green functions on the three
dimensional black
holes. By making use of them, we obtained free energies and entropies.
The divergent terms of
the entropies, which come from short distances,
are proportional to $ \dH  = \sqrt{\rp^2 -\rmi^2} $. Thus the Bekenstein-
Hawking formula is not satisfied except for the $ J = 0 $ case.
( For $ J \neq 0 $, there is another divergence which is the same as that of
the free energies. ) Furthermore, the divergences are not always
due to the existence of
the outer horizon and depend upon the regularization method.
In addition, the results obtained in approach I and II are quite
{\it different}.

Therefore we conclude that the Bekenstein-Hawking formula including
the quantum scalar fields is not valid in general.
Thus the relationship between the divergence of the entropies and
the renormalization of the gravitational coupling constant is not clear.
It is also obscure whether the divergences are due to the existence of
the horizon.

The expressions of the entropies largely depend upon the method
of calculation, boundary conditions and regularization, namely, upon its
definition. Hence it is quite important to consider what kind of definition
we should adopt. Without fixing the definition, we cannot discuss
whether the entropies of scalar fields have
a meaning as the number of states or whether they can be regarded
as `` geometric entropy ''. Therefore we cannot understand the
problem of the relationship between information loss and entropy of quantum
fields until the problem of the definition is settled.

Fortunately, we have got explicit expressions of the thermodynamic quantities
of scalar fields on the three dimensional black holes. Thus we think that
it is possible to apply our results to various problems, for example,
the problems discussed above, Hawking radiation \cite{Ha} and
the generalized second law \cite{Be}.
Since the divergences appearing in our calculation can
be absorbed into the renormalization of the {\it cosmological constant },
\nobreak
it may worth investigating the relationship between the divergence of the
entropies and renormalization.
%
%%%%%%%%%%%%%%%%%%%
%\input{notesty}
%\input{lmydef}
%\input{bhdefs}
%\begin{document}
%%%%%%%%%%%%%%%%%%%%
%\parbigskipn
%\begin{center}
%{\Large\sc Acknowledgement}
%\end{center}
\parbigskipn
\csectionast{ Acknowledgement }
%%%%%%%%%%%%%%%%%%%%%%
We would like to thank K. Shiraishi for helpful answer to questions
about his works.
Y. S. also acknowledges discussions with
T. Yamamoto, T. Tani and T. Izubuchi.
The research of Y. S. is partially supported by the JSPS Research
Fellowship for Young Scientists ( No. 06-4391 ) from the Ministry
of Education, Science and Culture.

%%%%%%%%%%%%%%%%%%%%
\newpage
%\parbigskipn
%\begin{center}
%{\Large\sc Appendix}
%\end{center}
\csectionast{ Appendix }
\setcounter{equation}{0}
         \renewcommand{\theequation}{A .\arabic{equation}}
%%%%%%%%%%%%%%%%%%%%%%

In this appendix, we summarize the derivation of the Feynman Green function
in the universal covering space of three dimensional anti-de Sitter space
(CAdS${}_3) $. Quantization of a scalar field in CAdS${}_n$ is discussed
in \cite{AIS}-\cite{BL},
and the Feynman Green function is given \cite{BL}
in terms of hypergeometric functions. In
the three dimensional case, we find that the Feynman Green function is
simplified and expressed in terms of elementary functions.

We parametrize CAdS${}_3 $ as follows
\eqabegin
&& -u^2 - v^2 + x^2 + y^2 = - l^{-2} \comma \nn \\
&& u = l \sin \tau \sec \rho \comma \qquad
   v = l \cos \tau \sec \rho \comma \nn \\
&& x = l \sin \theta \tan \rho \comma \qquad
   y = l \cos \theta \tan \rho \comma
\eqaend
where $ 0 \leq \rho < \pi/2 \comma  \quad 0 \leq \theta <  2 \pi
\comma \quad  -\infty \ < \tau < \infty $.
Then the metric becomes
\eqabegin
  d s^2 &=&  l^2 \sec^2 \rho \lbra  - \ d \tau^2 + d \rho^2
                 + \sin^2 \rho \ d \theta^2 \rbra
                 \period
\eqaend
The field equation for a scalar field is given by
\eqabegin
 \lbra  \Box - \mu l^{-2} \rbra \psi (x) &=& 0 \period
\eqaend
Making the separation of variables
\eqabegin
 \psi_{m \omega} &=& \ e^{- i \omega \tau} \ e^{i m \theta} R(\rho)
  \comma \qquad ( m \in {\bf Z} ) \comma
\eqaend
the equation for the radial function $ R(\rho)$ is given as
\eqabegin
 \lbra \del_\rho^2 + \frac{1}{\sin \rho \ \cos \rho} \del_\rho
  + \omega^2 - \frac{m^2}{\sin^2 \rho} - \mu \sec^2 \rho \rbra
   R(\rho) &=& 0
  \period
\eqaend
Let us make the change of variables
$ v = \sin^2 \rho \comma $
and define a function $ f(v) $ by
\eqabegin
 R(\rho) &=& v^{\abs{m} /2} ( 1-v )^{\lambda/2} f(v) \comma
\eqaend
with
\eqabegin
 \lambda &=& \lambda_\pm \, \equiv \, 1 \pm \sqrt{1 + \mu}
 \period
\eqaend
Then the radial equation above is reduced to the hypergeometric equation
\eqabegin
 \lbbra v(1-v) \del_v^2 + \lmbra c - ( a + b + 1 )v \rmbra \del_v
  -ab \rbbra \ f(v) &=& 0
  \comma
\eqaend
where
\eqabegin
 a &=& \half ( \lambda + \abs{m} - \omega ) \comma \nn \\
 b &=& \half ( \lambda + \abs{m} + \omega ) \comma \nn \\
 c &=& \abs{m} + 1 \period
\eqaend
If we require the regularity at $ v = 0 $, the solution is expressed by
the Gauss' hypergeometric function $ F $ as
\eqabegin
 f(v) &=& F(a, b; c; v)
 \period
\eqaend

Since  CAdS${}_3$ is not globally hyperbolic, it is necessary to
impose boundary conditions at the spatial infinity. Following
\cite{AIS}-\cite{MT},
we require the condition to conserve energy.
This means that the surface integral
of the energy-momentum tensor at the spatial infinity must vanish. This
requirement leads to
\eqabegin
 \abs{\omega} &=& \lambda + \abs{m} + 2 n \qquad ( n = 0, 1, 2, .., )
 \comma
\eqaend
where
\eqabegin
 \lambda &=& \lmbra
    \begin{array}{ll}
       \lambda_\pm & {\rm for} \ 0 > \mu > -1 \comma  \\
       \lambda_+   & {\rm for} \ \mu \geq 0 \comma  \mu = -1
    \end{array}
    \right.
    \period
\eqaend
Then $ a $ becomes zero or a negative integer.
Thus by using a mathematical formula
\cite{AS} we get
\eqabegin
 \psi (x) &=& \sum_{m, n} \lbbra a_{mn} \psi_{m n} +
 ( a_{mn} \psi_{m n} )^{\ast} \rbbra
  \qquad ( m \in {\bf Z} , \quad n = 0, 1, 2, ... ) \comma \\
 \psi_{m n} &=& C_{mn} \ e^{- i \omega \tau} \ e^{ i m \theta}
      \lbra \sin \rho \rbra^{\abs{m} } \lbra \cos \rho \rbra^{\lambda}
      P_n^{(\abs{m} , \lambda-1)} ( \cos 2 \rho )
      \label{psimn} \comma
\eqaend
where $ P_n^{(\alpha,\beta)} $ is a Jacobi Polynomial and $ C_{mn} $
is a normalization constant.

For the positive frequency part $ \psi^{(+)} $ of the solution
we can define a positive definite scalar product as
\eqabegin
 \lbra  \psi_1^{(+)} , \psi_2^{(+)} \rbra
 &\equiv& -i \int_{\Sigma} d^2 x \sqrt{-g} g^{0\nu}
       \psi_1^{(+)\ast} \mathop{\del_\nu}^{\leftrightarrow} \psi_2^{(+)}
         \comma
\eqaend
where $ \Sigma $ is a spacelike hypersurface.
Then the normalization constant $ C_{mn} $ is determined by the condition
$ \lbra  \psi_{mn}^{(+)} , \psi_{m'n'}^{(+)} \rbra =
\delta_{m m'} \delta_{n n'} $.
By using the orthogonal relation with respect to Jacobi Polynomials
\cite{AS},
\eqabegin
&& \int_0^{\pi/2} d \rho \tan \rho
      \lbra \sin \rho \rbra^{2 \abs{m} }
      \lbra \cos \rho \rbra^{2\lambda}
      P_n^{(\abs{m} , \lambda-1)}(\cos 2 \rho)
      \ P_{n'}^{(\abs{m} , \lambda-1) }(\cos 2 \rho) \nn \\
&=& \delta_{n n'}
     \frac{1}{2(2n \splus \lambda \splus \abs{m} ) }
        \frac{\Gamma(n \! + \! \abs{m} \! + \! 1)\Gamma(n \!+ \! \lambda)}
             {n! \Gamma( n \splus \lambda \splus \abs{m} )}
 \comma \label{orthoP}
\eqaend
we get
\eqabegin
 C_{mn} &=& \lbbra \frac{n! \ \Gamma(\abs{m} \splus \lambda \splus n )}
              {2 \pi l ( \abs{m} \splus n )! \ \Gamma(\lambda \splus n ) }
              \rbbra^{1/2}
              \period
\eqaend

Now we quantize the scalar field by setting the commutation relation
\eqabegin
 \lbbra a_{mn}, a_{m'n'}^\dagger \rbbra = \delta_{mm'} \delta_{nn'}
      \period
\eqaend
Then we get
\eqabegin
 \lbbra \psi(x), \ \psi(x') \rbbra_{\tau=\tau'} &=& 0 \comma \nn \\
 \lbbra \psi(x), \ \del_{\tau'} \psi(x') \rbbra_{\tau=\tau'} &=&
 -i \frac{1}{g^{\tau\tau} \sqrt{-g}} \delta(\theta-\theta')
    \delta(\rho - \rho')
  \period \label{comm}
\eqaend
Here we have used the orthogonal relation (\ref{orthoP}). The $ \delta $
function is defined for the space of functions of the form as (\ref{psimn})

Let us define
\eqabegin
 -i \Gf(x,x') &=& \bra{0} T \lmbra \psi(x) \psi(x') \rmbra \ket{0} \nn \\
 &\equiv& \theta(\tau -\tau') \sum_{m \comma  n}
  \psi_{mn}(x) \psi_{mn}^{\ast} (x')
         + \ ( x \leftrightarrow x' )
  \period
\eqaend
{}From (\ref{comm}), it follows that
\eqabegin
 \lbra \Box - \mu l^{-2} \rbra \Gf(x, x') &=&  \frac{1}{\sqrt{-g}}
         \delta (x - x' ) \comma
\eqaend
namely $ \Gf $ is the Feynman Green function.

Furthermore, we can perform the summation with respect $ m $ and $ n $.
First, we can set $ x' = (\tau', \rho' , \theta') = (0,0,0) $ , ( i.e.
$ (u',v', x',y') = (0,l,0,0) $) without loss of generality because
 CAdS${}_3 $ is homogeneous. Then only the term with $ m = 0 $ contribute
to the summation, i.e.,
\eqabegin
 -i \Gf (x,0) &=& \frac{1}{2\pi l} \ e^{-i\lambda \abs{\tau} }
  ( \cos \rho )^{\lambda}
   \sum_{n= 0}^{\infty} \ e^{-2in\abs{\tau} }
   \ P_n^{(0, \lambda-1)}( \cos 2 \rho)
   \period
\eqaend
By making use of the mathematical formulae \cite{PBM}
\eqabegin
&& \sum_{k = 0}^{\infty} \frac{(\alpha \splus \beta \splus 1)_k}
       {(\beta \splus 1)_k} \ t^k \ P_k^{(\alpha, \beta)}(x) \nn \\
  && \qquad = ( 1 + t )^{-\alpha -\beta -1}
   F \lbra \frac{\alpha \splus \beta \splus 1}{2},
     \frac{\alpha \splus \beta \splus 2}{2}; \beta + 1 ;
     \frac{2 t (x \splus 1)}{(t \splus 1)^2}
   \rbra
   \comma
\eqaend
we get
\eqabegin
 - i \Gf(x,0) \equiv - i \Gf(z) &=&
           \frac{l^{-1}}{2^{\lambda+1} \pi} z^{-\lambda} \
                  F \lbra \half \lambda , \half (\lambda \splus 1) ;
                  \lambda ; z^{-2} \rbra
                     \period
\eqaend
Here $ z $ is defined by
\eqabegin
 z &=& 1 + l^{-2} \sigma (x, 0) + i \varepsilon \comma \label{epGf}
\eqaend
and
$ \sigma(x,x') $ is half of the distance between $ x $ and $ x'$ in the four
dimensional embedding space, namely,
\eqabegin
  \sigma (x, x') &=& \half \eta_{\mu\nu} ( \xi - \xi')^{\mu}( \xi - \xi')^{\nu}
              \comma
\eqaend
where $ \eta_{\mu\nu} = diag( -1,-1, +1,+1 )$ and $ \xi $ and $ \xi' $ are
the coordinates in the embedding space. The infinitesimal imaginary part
$ i \varepsilon \quad ( \varepsilon > 0 ) $
in $ z $ is added so that the Green function looks locally like the Minkowski
one \cite{AIS}.  In the three dimensional case, from the mathematical
formula,
\eqabegin
F\lbra  a, \half \splus a ; 2 a ; z \rbra &=&
   2^{2a-1} (1-z)^{-1/2} \lbbra  1 + ( 1-z)^{1/2} \rbbra^{1-2a}
   \comma
\eqaend
we find that the Feynman Green function is simplified to be
\eqabegin
     - i \Gf(z) &=& \frac{l^{-1}}{4 \pi} ( z^2 -1)^{-1/2}
                \lbbra  z + (z^2-1)^{1/2} \rbbra^{1-\lambda}
         \period
\eqaend
This result is obtained also by replacing $ \abs{\tau} $
with $ \abs{\tau} - i \varepsilon $ so that $ \abs{e^{-2in\abs{\tau} } }
< 1 $ holds and by utilizing the generating function of
Jacobi Polynomials.

In the case with general $ x' $ , we have only to replace $ \sigma(x, 0) $
with $\sigma(x, x') $.

%\end{document}
%%%%%%%%%%%%%%%%%
\newpage
%%%%%%%%%%%%%%%%%
\def\thebibliography#1{\list
 {[\arabic{enumi}]}{\settowidth\labelwidth{[#1]}\leftmargin\labelwidth
 \advance\leftmargin\labelsep
 \usecounter{enumi}}
 \def\newblock{\hskip .11em plus .33em minus .07em}
 \sloppy\clubpenalty4000\widowpenalty4000
 \sfcode`\.=1000\relax}
\let\endthebibliography=\endlist

\csectionast{ References }

%%%%%%%%%%%%%%%
\newpage
%%%%%%%%%%%%%%%%%%

%\vspace*{2.0cm}
\begin{center}
    \begin{minipage}[t]{4.0in}
          \epsfxsize = 4.0in
          \epsfbox{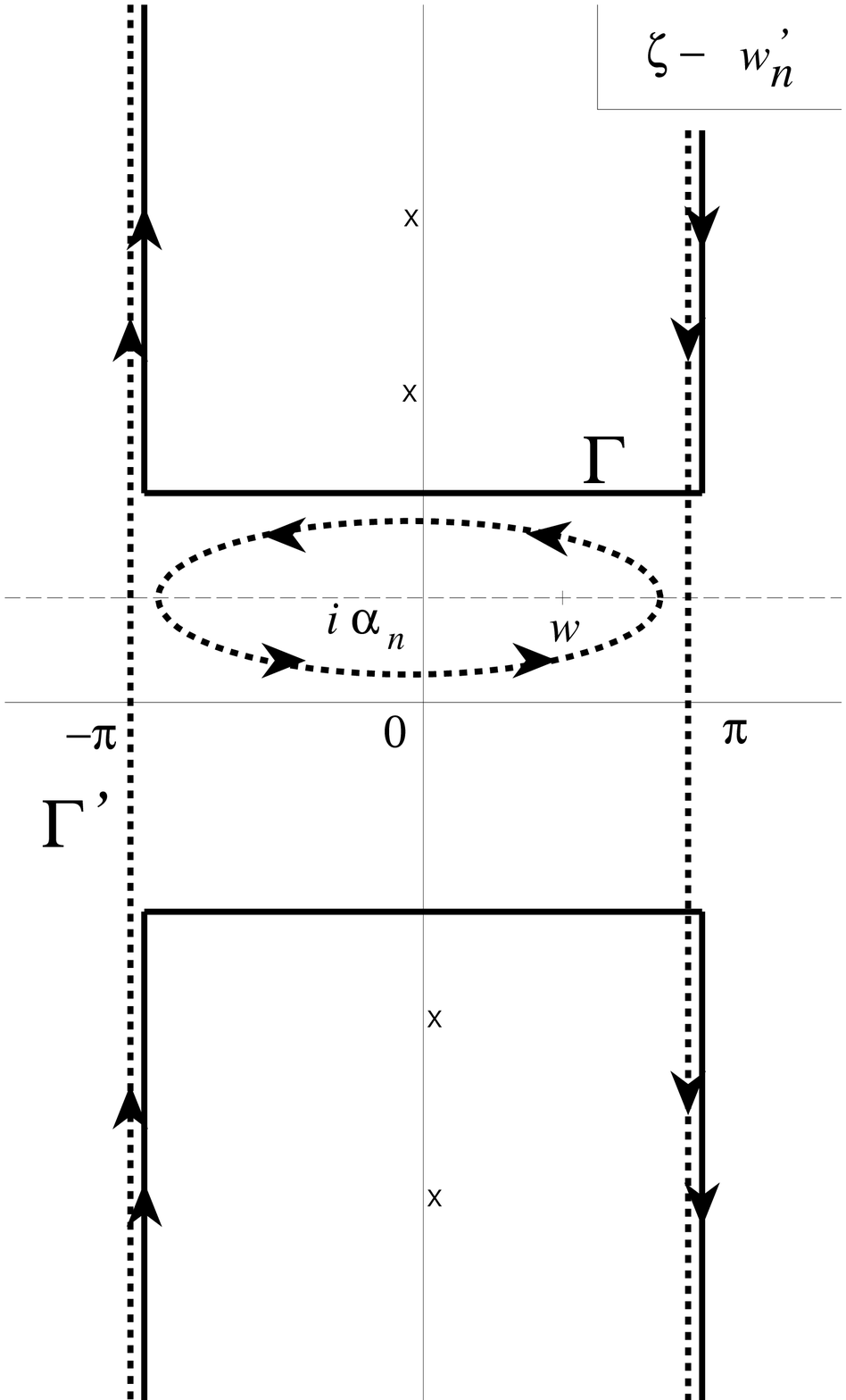}
   \end{minipage}
\end{center}

%\begin{center}
%  {\large\sc Figure Caption}
%end{center}
\vspace*{1.0cm}
\begin{description}
  \item[Fig. 1 \ :]
      Contour $ \Gamma $ ( solid line ) and Contour $ \Gamma' $
      ( dashed line ) in $ \zeta - w_n' $ plane. The crosses $ (\times) $
      indicate the singularities of $ \tGf(\zeta - w_n')$ in the region
      $ - \pi < {\rm Re } \ $
       $(\zeta - w_n') \leq \pi $ for $ r,\ r' \geq \rp $.
      $ \alpha_n = \rmi \delphi^+_n /l $ . In this figure, we
      show the contour $ \Gamma $ in the case of small $ \abs{\alpha_n} $.
      In the case of
      large and positive $ \alpha_n $, for example,
      the line $ \zeta - w_n' = i \alpha_n $ is above the crosses, and
      we can not take a contour as $ \Gamma $ in this figure. In this
      case we have only to deform $ \Gamma $  maintaining
       the property that  it can be deformed into a
      contour topologically equivalent to $ \Gamma' $.
 \end{description}

%%%%%%%%%%%%%%%
\end{document}